\documentclass[11pt]{article}
\usepackage{jheppubmodnew}
\pdfoutput=1
\usepackage{hyperref}
\usepackage{psfrag}
\usepackage{array}
\usepackage{amssymb}
\usepackage{amsmath}
\usepackage{amsthm}
\usepackage{mathtools}
\usepackage{graphicx}
\usepackage[labelsep=quad]{subcaption}
\usepackage{cite}
	
\usepackage[punctsep]{collref}
\collectsep[]{;}
\usepackage{epsfig}
\usepackage{tikz}
\usetikzlibrary{arrows,plotmarks,positioning,calc}
\newcommand{\alset}{{\cal A}}

\newtheorem*{definition}{Definition}
\newtheorem*{observation}{Observation}
\newtheorem*{puzzle}{Puzzle}

\def\pbound{P_{B}}
\def\hmat{\rho}

\def\al{A}

\def\pb[#1,#2]{\{#1, #2\}}
\def\deb[#1,#2]{[#1,#2]_{\text{D.B.}}}

\def\tr{{\rm Tr}}

\def\Or[#1]{{\text{O}}\left({#1}\right)}
\def\dotl[#1,#2]{\left\langle #1,\, #2 \right\rangle}
\def\dotlb[#1,#2]{\left\langle #1,\, #2 \right\rangle}
\def\dotlm[#1,#2]{\left[ #1,\, #2 \right]}
\def\dotp[#1,#2]{(\vect{#1} \cdot\vect{#2})}
\def\aff[#1,#2]{\hat{#1}(#2)}

\def\n4sym{{\cal N}=4 SYM}
\def\>{\rangle}
\def\<{\langle}
\def\weight[#1,#2,#3]{\{(#1),#2,#3\}}
\def\ads[#1]{$\text{AdS}_{#1}$}

\newcommand{\be}{\begin{equation}}
\newcommand{\ee}{\end{equation}}
\newcommand{\ba}{\begin{align}}
\newcommand{\ea}{\end{align}}
\newcommand{\bs}{\begin{split}}
\def\sess\end{split}
\newcommand{\vect}[1]{{\boldsymbol{#1}}}

\def \bea {\begin{eqnarray}}
\def \eea {\end{eqnarray}}
\def \bea* {\begin{eqnarray*}}
\def \eea* {\end{eqnarray*}}

\def \bes {\begin{equation*}}
\def \ees {\end{equation*}}

\def \wr {\varrho}
\newcommand{\dt}{\tilde{d}}
\def\alcut[#1]{{\cal A}_{#1, \epsilon}}
\def\alseg[#1,#2]{{\cal B}_{#1, #2}}

\def\supcharge[#1]{\{#1\}}

\def\projsupeig[#1]{{\cal P}_{{\ell, m}}[{#1}]}
\def\transop[#1, #2]{T_{\{#1\}, \{#2\}}}
\def\supket[#1]{|\{#1\} \rangle}
\def\supbra[#1]{\langle \{#1\} | }

\newcommand{\htt}{h^{\text{TT}}}
\newcommand{\htr}{h^{\text{T}}}
\newcommand{\hlo}{h^{\text{L}}}

\title{Inconsistency of Islands in Theories with Long-Range Gravity}
\author{Hao Geng$^{a,b}$, Andreas Karch$^{b,c}$, Carlos Perez-Pardavila$^{c}$, Suvrat Raju$^d$, Lisa Randall$^a$, Marcos Riojas$^{c}$ and Sanjit Shashi$^{c}$}
\affiliation{$^a$Harvard University, 17 Oxford St., Cambridge, MA, 02139, USA.}
\affiliation{$^b$Department of Physics, University of Washington, Seattle, WA, 98195-1560, USA.}
\affiliation{$^c$Theory Group, Department of Physics, University of Texas, Austin, TX 78712, USA.}
\affiliation{$^d$International Centre for Theoretical Sciences, Tata Institute of Fundamental Research, Shivakote, Bengaluru 560089, India.}

\emailAdd{gengphysics666@gmail.com, karcha@utexas.edu, cjp3247@utexas.edu}
\emailAdd{suvrat@icts.res.in, randall@g.harvard.edu}
\emailAdd{marcos.riojas@utexas.edu, sshashi@utexas.edu}

\abstract{In ordinary gravitational theories, any local bulk operator in an entanglement wedge is accompanied by a long-range gravitational dressing that extends to the asymptotic part of the wedge. Islands are the only known examples of entanglement wedges that are disconnected from the asymptotic region of spacetime. In this paper, we show that the lack of an asymptotic region in islands creates a potential puzzle that involves the gravitational Gauss law, independently of whether or not there is a non-gravitational bath. In a theory with long-range gravity, the energy of an excitation localized to the island can be detected from outside the island, in contradiction with the principle that operators in an entanglement wedge should commute with operators from its complement. In several known examples, we show that this tension is resolved because islands appear in conjunction with a massive graviton. We also derive some additional consistency conditions that must be obeyed by islands in decoupled systems. Our arguments suggest that islands might not constitute consistent entanglement wedges in standard theories of massless gravity where the Gauss law applies.}
\begin{document}
\maketitle
\section{Introduction \label{secintro}}
Recent literature \cite{Penington:2019npb,Almheiri:2019hni,Almheiri:2019psy,Almheiri:2019qdq,Almheiri:2019psf,Penington:2019kki} has presented significant progress in understanding the evaporation of AdS black holes coupled to an auxiliary non-gravitational bath. In these settings, the fine-grained entropy of a part of the bath, called the radiation region, can be computed through an elegant ``island rule.'' It is further believed that operators that are localized within the island can be reconstructed from operators in the radiation region in the same sense that, in standard AdS/CFT, operators from part of a boundary of AdS can be used to reconstruct operators in the corresponding entanglement wedge. Using these techniques, the entropy of the radiation region has been found to follow a Page curve.

In spacetime dimensions larger than two, precise computations of the Page curve have been performed using a doubly-holographic setup where the AdS black hole and non-gravitational bath are realized through a Karch-Randall brane \cite{Karch:2000ct,Karch:2000gx} embedded in a higher-dimensional AdS spacetime \cite{Almheiri:2019psy}. (See \cite{Rozali:2019day,Liu:2020gnp,Sully:2020pza,Emparan:2020znc,Liu:2020jsv,Ling:2020laa,KumarBasak:2020ams, Caceres:2020jcn,Caceres:2021fuw,Deng:2020ent,Karlsson:2021vlh,Miao:2021ual,Bachas:2021fqo,May:2021zyu,Kawabata:2021hac,Bhattacharya:2021jrn,Kim:2021gzd,Aalsma:2021bit,Neuenfeld:2021wbl,Geng:2021iyq,Balasubramanian:2021wgd,Uhlemann:2021nhu,Neuenfeld:2021bsb,Kawabata:2021vyo,Chu:2021gdb,Kruthoff:2021vgv,Akal:2021foz,KumarBasak:2021rrx,Lu:2021gmv,Omiya:2021olc, Manu:2020tty}
for recent related work.) In this setting, it was pointed out in \cite{Geng:2020qvw} that the lower-dimensional graviton is always massive.\footnote{Specifically, there is a tower of gravitational KK modes whose lightest graviton is massive.} This is a manifestation of a more general phenomenon: when a gravitational theory in AdS is coupled to a non-gravitational bath, the graviton in AdS picks up a mass. Nevertheless, it is sometimes believed that the non-gravitational bath and the massive graviton that appear in such models are merely technicalities, and that the general lessons regarding islands and the Page curve should be applicable to other physical systems including realistic black holes in asymptotically flat space \cite{Krishnan:2020fer,Almheiri:2020cfm}.

However, in previous work \cite{Geng:2020fxl}, we pointed out that the non-gravitational bath is not just a spectator but an important participant in the physics. When gravity is dynamical in the bath, as it should be in realistic models of black holes, it was found that the fine-grained entropy of radiation was constant, consistent with the previously obtained results \cite{Laddha:2020kvp} that the Page curve of radiation is trivial for black holes in asymptotically flat spaces.\footnote{Even in the presence of dynamical gravity, the Page curve may be the answer to appropriate non-gravitational questions \cite{Geng:2020fxl}. Also see \cite{Ghosh:2021axl,Geng:2021wcq,Langhoff:2020jqa} and section 4.2 of \cite{Laddha:2020kvp} for a discussion of whether coarse-graining the entropy of the radiation in the presence of dynamical gravity may lead to a nontrivial Page curve.}

In this paper, we return to the system with a non-gravitational bath and present an argument that suggests the mass of the graviton plays a significant physical role in allowing islands to constitute an entanglement wedge.

The crux of our argument is very simple. In ordinary gravitational theories, there are no local gauge-invariant operators. We note that if one is studying a non-gravitational observable where gravity would be a small perturbation, it is possible to define approximately local observables by choosing gauge  that would suffice for such a measurement.  However, when studying processes where both gravitational and quantum effects are important, there is no procedure for defining a local observable, perturbatively or otherwise. 

We now restrict our attention to theories with gravity. As we review below, in standard gravitational theories with massless gravitons, every ``localized'' operator must be dressed in some way to the asymptotic boundary. This dressing is sometimes referred to as a gravitational Wilson line, which terminates at the asymptotic boundary. In ordinary examples of entanglement wedges in AdS/CFT \cite{Maldacena:1997re,Gubser:1998bc,Witten:1998qj}, as we review in section \ref{secasymptotic}, the connected components of the wedge contain a piece of the asymptotic boundary. So one can meaningfully localize operators from such a wedge by dressing them to this part of the boundary. These operators commute with operators from the complement of the wedge when the latter are dressed to the complementary asymptotic region.

However, an island represents a unique type of entanglement wedge that is entirely surrounded by its complement and where the wedge itself does not extend to the asymptotic boundary. When a region is surrounded by its complement, and it is the complement that extends to the asymptotic boundary, it was argued in \cite{Laddha:2020kvp,Chowdhury:2020hse} based on a careful analysis of the gravitational constraints that the state of the region could be completely determined through observations in its complement. A review of this result, termed the ``principle of holography of information,'' can be found in \cite{Raju:2020smc}. It is apparent that the picture of islands is already in tension with this principle. Nevertheless, in this paper, we will {\em not} need to invoke the full power of the principle of holography of information. We will demonstrate that 
simple physical principles suffice to generate the following puzzle for islands.

Consider a simple unitary operator that adds a localized excitation to the island. Since the excitation is confined to the island, which has finite extent, it must have some nonzero energy by the Heisenberg uncertainty principle. In theories with massless gravitons, the energy can be measured from the falloff of the asymptotic metric using the Gauss law. This would imply that any unitary operator that creates an excitation in the island must fail to commute with the metric in the complement of the island. This is inconsistent with the idea that the algebra of an entanglement wedge should be closed and commute with the algebra of its complement.

This puzzle becomes even more acute if the island under consideration is described by operators from a radiation region in a non-gravitational system, the bath, and its complement is described by operators from the complement of the radiation region. In this setting, operators that act on the island must commute with operators that act on its complement since, in the non-gravitational theory, operators in the radiation region and its complement are spacelike to each other and so commute by microcausality. 

In this paper, we argue that one way to address this puzzle is with massive gravity.  In fact, the mass of the graviton appears naturally when gravity in AdS is coupled to an external bath \cite{Aharony:2003qf}. The reason is simply that, since the stress-tensor on the boundary of AdS is no longer conserved, it picks up an anomalous dimension. This corresponds to a nonzero mass for the graviton in AdS. 

\begin{figure}[!ht]
\begin{center}
\includegraphics[height=0.3\textheight]{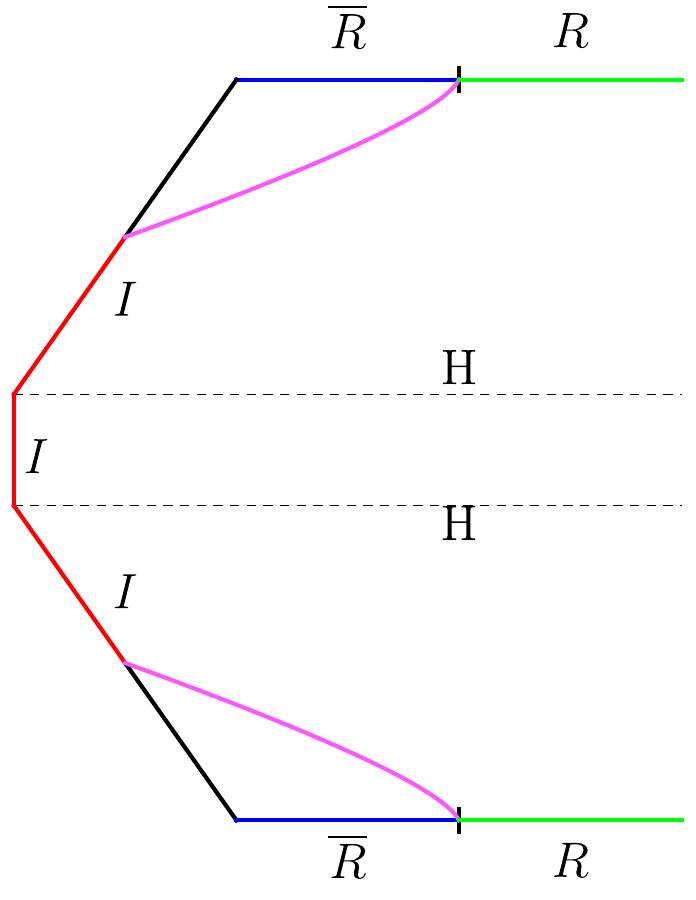}
\caption{\em A cartoon of a constant-time slice of a black hole with a brane embedded. $R$ is the union of regions on two asymptotic boundaries and $\overline{R}$ is its complementary region. The horizons in the bulk are marked by $H$. The separation between horizons is meant to convey that the Cauchy slice under examination is a late-time slice 
on which the wormhole is of a finite length. The dominant RT surface for the region $R$ is shown in purple. The region on the brane marked $I$ becomes the ``island'' in the lower-dimensional picture of Figure \ref{figlowerd}. In this figure, both the horizontal and the vertical directions are spatial. \label{fighigherd}} 
\end{center}
\end{figure}

This can be studied in the Karch-Randall scenario 
that is commonly used to study islands in $d > 2$ dimensions. Here, as mentioned above, an AdS$_d$ brane is embedded in an AdS$_{d+1}$ black hole spacetime. The radiation region $R$ is just a part of the non-gravitational boundary of this AdS$_{d+1}$. The entropy of $R$ can be computed using the standard RT/HRT prescription \cite{Ryu:2006bv,Ryu:2006ef,Hubeny:2007xt} and is given by the area of a bulk minimal surface. In some cases, this surface may end on the brane as shown in Figure \ref{fighigherd}. 

\begin{figure}[!ht]
\begin{center}
\includegraphics[height=0.3\textheight]{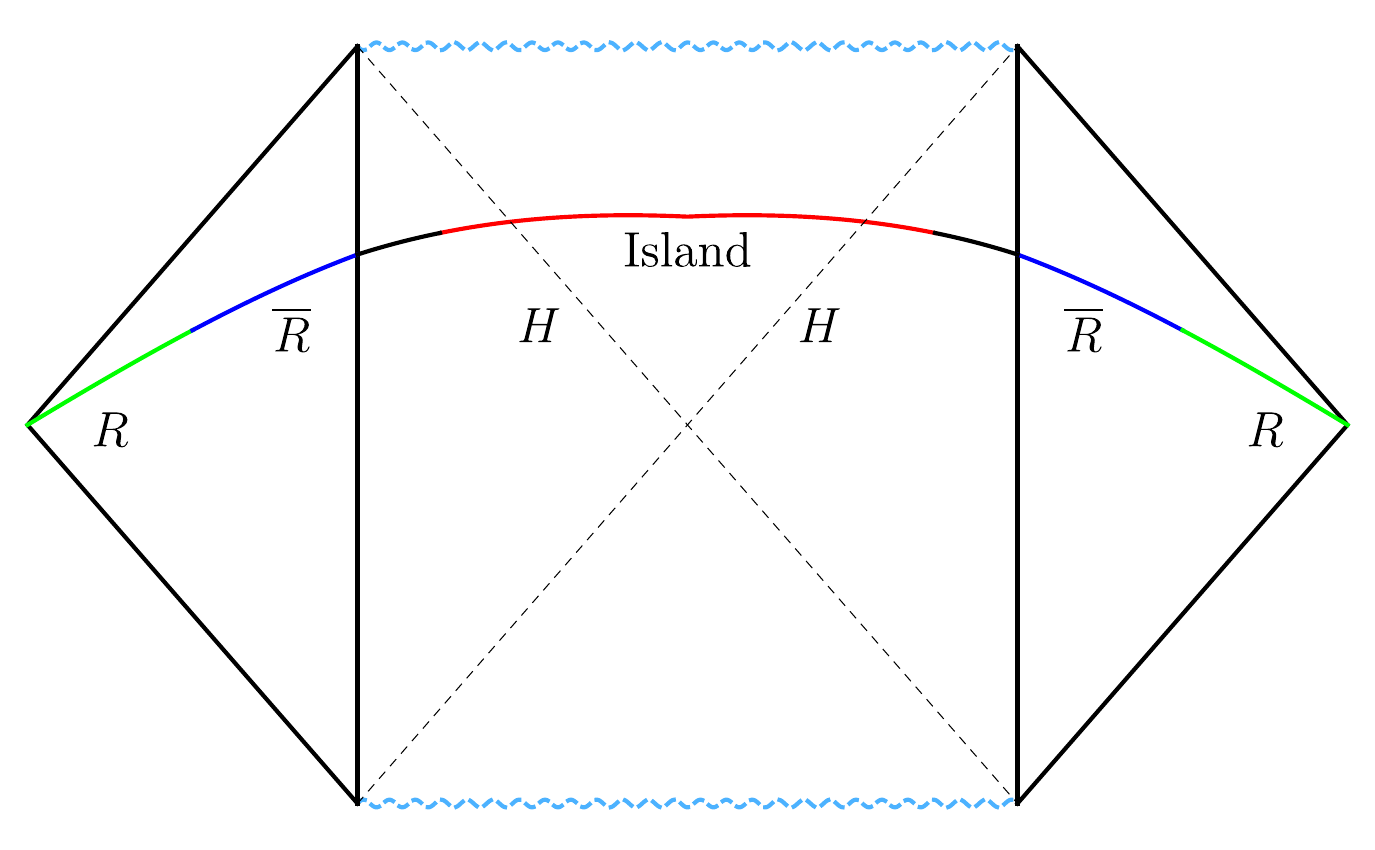}
\caption{\em A spacetime diagram of the same system of branes in a black hole in the $d$ dimensional description. The entanglement wedge for the region $R$ is now an ``island''. In this figure, the horizontal direction is spatial and time runs along the vertical direction. \label{figlowerd}} 
\end{center}
\end{figure}

The entanglement wedge of the radiation region $R$ in Figure \ref{fighigherd} is a conventional entanglement wedge. This entanglement wedge takes on the form of an island if one uses the dual $d$-dimensional description to obtain a black hole in AdS$_d$ coupled to a non-gravitational bath as shown in Figure \ref{figlowerd}. A striking aspect of this $d$ dimensional description with a non-gravitational bath is that although there is a localized graviton that ``locally'' generates a lower-dimensional gravitational theory, this lower-dimensional graviton is massive
 \cite{Karch:2000ct,Porrati:2003sa,Porrati:2001gx,Porrati:2002dt}. The constraints in massive gravity are significantly weaker than in massless gravity and do {\em not} disallow localized excitations, thereby resolving the puzzle presented above. 

This is why both pictures of Figure \ref{fighigherd} and Figure \ref{figlowerd} are consistent. There is no puzzle in Figure \ref{fighigherd} because the entanglement wedge extends to the asymptotic region and there is no ``island'' in the entanglement wedge, i.e. the entanglement wedge has no connected component that is separated from the boundary. There is an ``island'' in Figure \ref{figlowerd}, but the graviton is massive. 

This discussion suggests that the paradigm of islands is inapplicable to standard theories of gravity with massless gravitons.

\subsection{Definitions and clarifications}\label{definition}

In this paper we will use the phrase ``island'' strictly in accordance with the following definition.
\begin{definition}
An island is an entanglement wedge in the gravitating spacetime that does not extend to the asymptotic boundary of the gravitating spacetime.
\end{definition}
Accordingly, when ``islands'' are studied by embedding a brane in a higher-dimensional theory, we use the term ``island'' only in the lower dimensional dual description. We do {\em not} use the term ``island'' to describe the full higher-dimensional entanglement wedge since that does extend to the boundary of the higher-dimensional spacetime. We urge the reader to keep this definition in mind for the rest of this paper, particularly since the term ``island'' is used more loosely in other parts of the literature.

We also clarify that we use the phrase ``entanglement wedge'' according to its original definition \cite{Headrick:2014cta} so that it is a region of the gravitating spacetime that can be reconstructed from some part of the non-gravitational spacetime. Some papers in the literature use the phrase ``entanglement wedge'' to include a part of the non-gravitational spacetime as well, but we will not use this convention. 
\paragraph{\bf Gauss law.}
When we refer to the {\em Gauss law} in this paper, we are referring to the relationship between the total energy on a Cauchy slice and the integral of an appropriate component of the asymptotic metric. This relationship follows from the Hamiltonian constraint in standard theories of gravity.  We emphasize that we will use the Gauss law not just as a relationship between expectation values but also inside quantum correlation functions. For clarity, we will always refer to the Hamiltonian constraint as a constraint equation to distinguish it from the Gauss law, as defined above. Note that even in massive gravity, states must obey local constraint equations, which we review below. However these constraints do not lead to a Gauss law.

\paragraph{Bulk reconstruction \\}
In this paper, we adopt the perspective that a consistent entanglement wedge is one where it is possible to reconstruct approximately local bulk operators that, in the limit $\ell_{\text{pl}} \rightarrow 0$, reduce to standard quantum-field operators. From its beginning \cite{Hamilton:2005ju} the bulk reconstruction program has sought to reconstruct such operators. So, our perspective aligns with the standard perspective on subregion duality \cite{Faulkner:2017vdd}. 

We note that unless one can understand local physics in the bulk from the boundary, it does not make sense to state that a boundary subregion is dual to a bulk subregion. Moreover, the idea that an entanglement wedge can be demarcated by a precisely defined quantum extremal surface presupposes that one can localize bulk operators in the wedge to sub-AdS scales. Finally we note that our perspective is consistent with every known example of subregion duality that has been studied in the literature.

In appendix \ref{appalternatives} we relax the criterion of strict locality. For the reasons outlined above, we do not consider proposals where the only operators that can be reconstructed in the island are infinitely delocalized or spread out over a parametrically large spacetime region. We do however examine  (subject to the above restriction) the reconstruction of multi-local operators in the island and show that the proposals suggested so far are also subject to our puzzle.

\paragraph{Localization of quantum information in quantum gravity \\}
The arguments in this paper do {\em not} contradict the idea that quantum information is localized very differently in theories of gravity than it is in quantum field theories. The conclusion of \cite{Laddha:2020kvp}, which is consistent with the results of this paper, can be interpreted as the claim that information about a black hole microstate is always available outside for sufficiently detailed measurements in a standard theory of long-range gravity. This could not possibly be true in a local quantum field theory, where spacelike-separated operators commute.

However, islands that lead to a Page curve involve a ``halfway'' description. The island picture suggests that while operators in what is called the ``radiation region'' can be used to reconstruct the island, they {\em cannot} be used to reconstruct the complement of the island which is described by a commuting set of operators. In this halfway description, the unusual localization of information in gravity
is important because the radiation region describes degrees of freedom that are in a region that is spacelike separated from it. Nevertheless, the Hilbert space still effectively factorizes as in a local quantum field theory. The puzzles that we present below suggest that this halfway description can be valid only under certain conditions.

Our focus in this paper is on higher-dimensional theories of gravity in which the notion of a ``graviton'' makes sense. We briefly comment on two-dimensional models in section \ref{secdiscussion}. We caution the reader that additional subtleties might arise in two-dimensional theories, and the analysis of section \ref{secpuzzle} and \ref{secmassiveresolve} is not directly applicable to models of islands in two dimensions.

This paper is organized as follows. In section \ref{secasymptotic}, we review conventional entanglement wedges in AdS/CFT and emphasize the importance of the asymptotic region. In section \ref{secpuzzle}, we explore the puzzle sketched above, which involves the tension between the Gauss law and the appearance of entanglement wedges that are disconnected from the boundary of the gravitating space time. In section \ref{secmassiveresolve}, we examine the form of the constraints in massive gravity and show how the puzzle is avoided in this setting. We also study islands in doubly holographic settings and show how a dimensional reduction of the higher-dimensional constraints leads to the constraints of a lower-dimensional {\em massive} theory. In section \ref{secjt}, we discuss some additional constraints that are important when islands emerge in decoupled pairs of systems.

\section{Asymptotic regions in entanglement wedges \label{secasymptotic}}

In this section, we recount some simple properties of entanglement wedges and the algebra of operators associated with them. Here, we focus on conventional entanglement wedges in AdS/CFT; we will turn to islands in later sections. Our objective is to emphasize the significance of the fact that conventional entanglement wedges always have an asymptotic region.

\begin{figure}[!ht]
\centering
\begin{subfigure}{0.4\textwidth}
\centering
\includegraphics[width=0.29\textheight]{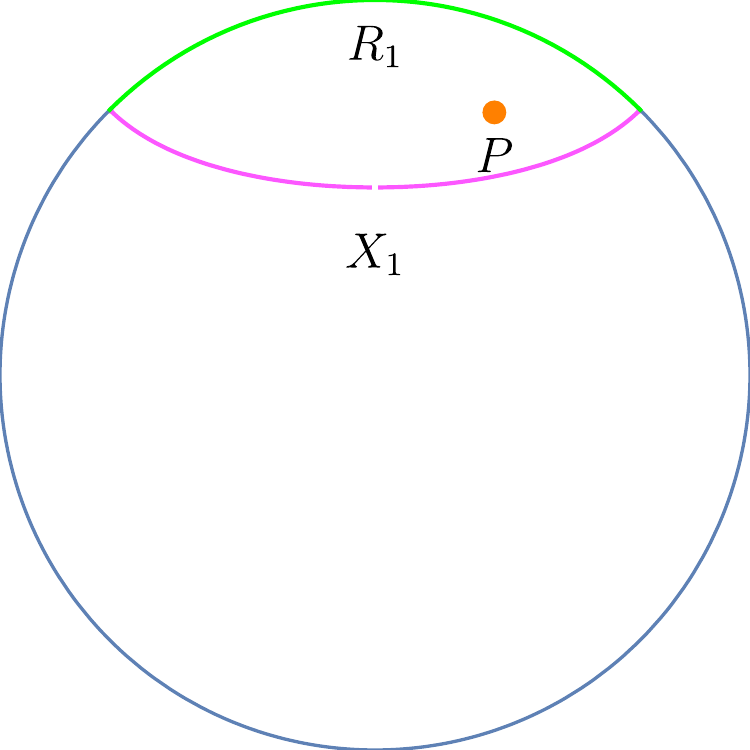}
\caption{\label{figwedgefirst}}
\end{subfigure}
\hspace{0.15\textwidth}
\begin{subfigure}{0.4\textwidth}
\centering
\includegraphics[width=0.29\textheight]{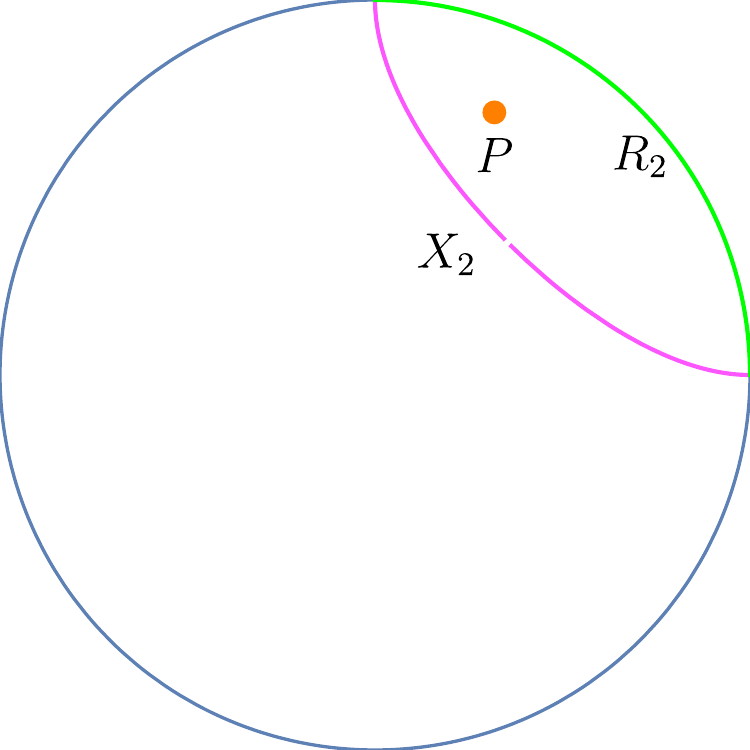}
\caption{\label{figentwedgesecond}}
\end{subfigure}
\caption{\em A point $P$ that is part of multiple entanglement wedges. The quasilocal bulk operator must be dressed to $R_1$ on the left and to $R_2$ on the right. The figure shows a time slice of global AdS. \label{figentwedge}}
\end{figure}

In a holographic theory, the entropy of a boundary region $R$ (see, for instance, Figure \ref{figentwedge}) is given by \cite{Ryu:2006bv,Ryu:2006ef,Hubeny:2007xt,Engelhardt:2014gca,Faulkner:2013ana},
\be
\label{holprescription}
S(R) = \text{min}\left[\text{ext}\left({A(X) \over 4 G} + S_{\text{bulk}}(E) \right) \right].
\ee

Here $X$ is a surface that is homologous to $R$, $E$ is the region bounded by $X$ and $R$
and $S_{\text{bulk}}(E)$ is the bulk entropy of the region $E$ computed while {\em ignoring} gravitational interactions. The homology constraint \cite{Headrick:2014cta,Headrick:2013zda} on $X$ states that $E$ should have {\em no other} boundaries except for $X$ and $R$. The bulk causal diamond constructed on the region $E$ is called the entanglement wedge of $R$. Since we will be interested in bulk and boundary regions on a single Cauchy slice, we simply refer to $E$ as the entanglement wedge and to its complement by $\overline{E}$. 

The ``subregion duality'' proposal \cite{Jafferis:2015del,Dong:2016eik} states that boundary operators in $R$ are dual to bulk operators in $E$. This can be made precise as follows. The boundary theory is non-gravitational and so one can associate an {\em algebra} of operators $\alset(R)$ with the region $R$. This algebra comprises all boundary operators that are localized within $R$ and so, by construction, it is closed under products, linear combinations, and Hermitian conjugation \cite{Haag:1992hx}. One can similarly associate an algebra with the complementary region, $\overline{R}$, and we denote this algebra by $\overline{\alset(R)}$.\footnote{When the boundary theory is a gauge theory, these algebras have a center \cite{Casini:2013rba,Ghosh:2015iwa}. However, this issue will be unimportant for the discussion in this paper.} Operators in $\alset(R)$ and $\overline{\alset(R)}$ commute,
\be
\label{commutcomplement}
[\al_1, \al_2] = 0, \quad \forall \al_1 \in \alset(R), \al_2 \in \overline{\alset(R)},
\ee
since every point in $R$ is separated by a spacelike interval from every point in $\overline{R}$. So equation \eqref{commutcomplement} follows from microcausality in the boundary theory. 
The subregion duality proposal is then that it is possible to find a representation of bulk operators associated with $E$ within $\alset(R)$ and a representation of bulk operators associated with $\overline{E}$ within $\overline{\alset(R)}$. 

The subregion duality proposal involves a subtle point that is sometimes glossed over. This aspect of the proposal can be illustrated by considering the case where the bulk theory has a gauge symmetry and charged matter. We caution the reader in advance that there are also important differences between gauge theories and gravitational theories that we will mention below, and so the gauge-theory discussion is provided only as a simplified warm-up.

When there is a gauge symmetry in the bulk, we expect to have a corresponding global symmetry in the boundary theory. We denote the generator of this global symmetry by $Q$ and we expect that it is given by the integral of a local boundary current $J$.
\be
Q = \int_{R} J + \int_{\overline{R}} J.
\ee

Consider a point in the entanglement wedge which we denote by $P$, and consider the operator that probes the charged matter field at the point $P$ which we denote by $\phi(P)$. By itself, $\phi(P)$ is clearly not gauge-invariant. One way to make it gauge-invariant is to attach a Wilson line $W(P, \pbound)$ to this operator that extends from $P$ in the bulk to another point $\pbound$ on the asymptotic boundary. We now have a gauge-invariant operator that, nevertheless, transforms nontrivially under the global charge.
\be
\label{qcomm}
[Q, W(P, \pbound) \phi(P)] = W(P, \pbound)\phi(P),
\ee
where we have normalized the charge to unity for simplicity. Note that although the operator transforms under the global charge, it is an allowed operator in the theory because it is invariant under ``small'' gauge transformations that die off at the asymptotic boundary. 

There is no unique choice of the path $P$ to $\pbound$ and not even of the boundary point $\pbound$ itself. But if we want to represent the operator $W(P,\pbound) \phi(P)$ as an element of $\alset(R)$, then the subregion duality proposal \eqref{commutcomplement} implies that it is necessary to choose $\pbound \in R$ and also ensure that the path between $P$ to $\pbound$ lies entirely within $E$. With this choice we have,
\be
\label{partcurrcommut}
\left[\int_{R} J,\,W(P,\pbound) \phi(P)\right] = W(P,\pbound) \phi(P); \qquad \left[\int_{\overline{R}} J,\,W(P,\pbound) \phi(P)\right] = 0,
\ee
in accordance with \eqref{commutcomplement}.

There are other options for making the operator $\phi(P)$ gauge-invariant. For example, one can simply fix the gauge. But the gauge-fixed operator must still obey \eqref{partcurrcommut} for it to be an element of $\alset(R)$. This leads to the following simple but robust conclusion.
\begin{observation}
A charged operator in an entanglement wedge can be represented as an operator in the dual boundary region only when it is dressed to that boundary region.
\end{observation}

We now turn to the analogous phenomenon in gravity. The difference between gauge theories and gravity is as follows. Gauge theories contain both positive and negative charges. Consequently, gauge theories contain an infinite number of exactly local gauge-invariant operators. An example of such an operator is a small Wilson loop that is entirely localized within a region. Similarly, in the example above, it is possible to construct a gauge-invariant localized operator entirely within an entanglement wedge by considering another operator of the opposite charge $\phi^*(\tilde{P})$ and connecting the two with a Wilson line: $\phi(P) W(P, \tilde{P} ) \phi^*(\tilde{P})$. But in gravity, there are no ``negative charges,'' so the gravitational dressing must extend to infinity and cannot terminate in the bulk.

A second way to understand the same physical fact is as follows. If an operator could be localized to a finite region, it would have nonzero energy just by the Heisenberg uncertainty principle. But since the energy can be measured near infinity in ordinary theories of gravity by the Gauss law, this operator cannot commute with the metric near infinity. This is a sign of the fact that even what may appear to be a ``local operator'' in gravity is secretly delocalized and must extend to the asymptotic boundary \cite{dewitt1960quantization, kuchar1991problem, Giddings:2005id,Jacobson:2012gh,Jacobson:2019gnm}.

In the context of subregion duality, if we want bulk operators in $E$ to be dual to operators in $R$ then they must be dressed to $R$. At leading order, this dressing can be described as follows. To make the operator $\phi(P)$ invariant under the gauge transformations of the theory, which comprise small diffeomorphisms---those that vanish near the asymptotic boundary---the position of the point $P$ is specified by relating
it to a part of the asymptotic boundary. In the semiclassical approximation this can be done by specifying the point $P$ to be the endpoint of a geodesic that starts at some point in the region $R$ and has a certain renormalized proper length. This picture is not very precise when fluctuations of the metric are themselves important. But, at an intuitive level, this relational prescription can be thought of as the analogue of a Wilson line that must be attached to charged local operators in gauge theories. 

Say that the operator $\phi(P)$ has been specified in a diffeomorphism-invariant manner as described above. Then the analogue of \eqref{qcomm} in gravity is that this operator transforms nontrivially under the boundary Hamiltonian. 
\be
\label{heisenbergeqn}
[H, \phi(P)] = -i {\partial \phi(P) \over \partial t}.
\ee

In this equation, the coordinate $t$ involves an extension of the boundary time coordinate into the bulk. By dressing the operator $\phi(P)$ in different ways, it is possible to choose different $t$ coordinates in the bulk and so the commutator of the boundary Hamiltonian with the bulk operator depends on the dressing. Note that the reason this is analogous to \eqref{qcomm} is that if we Fourier transform,
\be
\phi(P) = \int_{-\infty}^{\infty} \phi_{\omega} e^{i \omega t} d \omega, 
\ee
then \eqref{heisenbergeqn} tells us that the boundary Hamiltonian measures the energy of the Fourier components of $\phi(P)$,
\be
[H, \phi_{\omega}] = \omega \phi_{\omega}.
\ee

The boundary Hamiltonian can also be written as the integral of a local current density that is the sum of a term in $R$ and another in $\overline{R}$.
\be
H = \int_{R} T_{00} + \int_{\overline{R}} T_{00}.
\ee
Thus if we want the operator in the entanglement wedge $E$ to have a representation in the boundary region $R$, then the gravitational dressing must be chosen so that,
\be
\label{dressinggrav}
\left[\int_{R} T_{00},\,\phi(P)\right] = -i {\partial \phi(P) \over \partial t}; \qquad \left[\int_{\overline{R}} T_{00},\,\phi(P)\right] = 0.
\ee

The extrapolate dictionary \cite{Banks:1998dd} tells us that the boundary Hamiltonian is itself obtained as the limit of the bulk metric fluctuation in a certain gauge. Let $g_{\mu \nu}^{\text{AdS}}$ denote the background AdS metric with radial coordinate $r$.
If the bulk metric is expanded as $g_{\mu \nu}^{\text{AdS}} + h_{\mu \nu}$, where $h_{\mu \nu}$ is the deviation from AdS, then upon choosing Fefferman-Graham gauge near the boundary $r \to \infty$, i.e. $h_{r \mu} = 0$, the extrapolate dictionary reads \cite{deHaro:2000vlm},
\be
\label{extrapolatedict}
T_{0 0} = {d \over 16 \pi G} \lim_{r \rightarrow \infty} r^{d - 2} h_{00}.
\ee

We will provide a covariant version of this formula below. For now, we just note that the integral of equation \eqref{extrapolatedict} 
on the boundary of AdS provides the definition of the energy of the bulk state in a theory of gravity. Therefore \eqref{extrapolatedict} is just a manifestation of the Gauss law in the bulk since it tells us that the integral of the boundary metric fluctuation measures the energy of the state in the bulk.

The choice of dressing that ensures that equation \eqref{dressinggrav} holds also ensures that the operator $\phi(P)$ commutes with the metric fluctuation near the boundary in the region $\overline{E}$. This is consistent with the idea that operators in an entanglement wedge should commute with operators in its complement.

We can summarize this discussion in terms of the following observation.
\begin{observation}
Any bulk operator in an ordinary gravitational theory must be dressed to asymptotic infinity to make it invariant under small diffeomorphisms. The asymptotic part of an entanglement wedge provides a base that can be used to define relational observables in the bulk of the wedge.
\end{observation}

We should clarify that, for the purpose of this discussion, we need only the relationship \eqref{extrapolatedict} and the commutator \eqref{heisenbergeqn} to hold within low-point correlators, or the ``code subspace'' \cite{Almheiri:2014lwa}. The gravitational dressing is unimportant if we consider the limit where $G \rightarrow 0$, but it is already significant at leading nontrivial order in the gravitational constant. In fact, this nonzero commutator of bulk operators with the boundary Hamiltonian was emphasized when the concept of a code subspace was first introduced in the literature by considering small fluctuations about black hole microstates and termed the ``little Hilbert space'' in \cite{Papadodimas:2013jku}. In particular, even if \eqref{extrapolatedict} and \eqref{heisenbergeqn} receive corrections at higher orders in the gravitational constant or nonperturbative corrections, such corrections are not relevant for the discussion in this paper.

Second, this discussion has interesting consequences when we consider points that belong to two different entanglement wedges. In Figure \ref{figentwedge} we show a point $P$ that belongs to the entanglement wedge of the region $R_1$ and {\em also} to the entanglement wedge of the region $R_2$. When one implements the subregion duality proposal for region $R_1$, one picks an operator $\phi_1(P)$ that satisfies,
\be
\label{phionecomm}
\left[\int_{R_1} T_{00},\,\phi_1(P)\right] = -i {\partial \phi_1(P) \over \partial t}; \qquad \left[\int_{\overline{R_1}} T_{00},\,\phi_1(P)\right] = 0.
\ee

When one implements the subregion duality proposal for region $R_2$ one must pick an operator, $\phi_2(P)$, that satisfies
\be
\label{phitwocomm}
\left[\int_{R_2} T_{00},\,\phi_2(P)\right] = -i {\partial \phi_2(P) \over \partial t}; \qquad \left[\int_{\overline{R_2}} T_{00},\,\phi_2(P)\right] = 0.
\ee

Note that some operators may satisfy both equations \eqref{phionecomm} and \eqref{phitwocomm} but the point $P$ belongs to an infinite number of entanglement wedges and it is not possible to find a single operator that can serve as the bulk dual in all entanglement wedges. So it is important that one has some freedom in how to dress the bulk operator, and this freedom can be used to ``move around'' the commutator of the bulk operator with the asymptotic metric and the boundary stress tensor in order to ensure consistency with the subregion duality proposal. Nevertheless, whatever choice one makes for the dressing, there is always a nonzero commutator with the boundary Hamiltonian that is the integral of a component of the asymptotic metric on the entire boundary. In the example above this can be seen from the fact that both equations \eqref{phionecomm} and \eqref{phitwocomm} lead to equation \eqref{heisenbergeqn}.

\section{A puzzle with islands \label{secpuzzle}}

We now turn to islands and describe our puzzle. Islands can be understood as follows. We consider a CFT$_{\dt+1}$ propagating in a gravitating AdS$_{\dt+1}$ geometry. We couple this gravitational system to another system where the CFT$_{\dt+1}$ is supported in a non-gravitational spacetime. The coupling is designed to lead to ``transparent boundary conditions'' so that excitations in the CFT can propagate freely from the gravitating to the non-gravitational system.\footnote{We use $\dt$ in this section since in section \ref{secmassiveresolve} we will study islands that are realized on an AdS$_d$ brane embedded in AdS$_{d+1}$, in which case $\dt=d-1$.} The island rule \cite{Penington:2019npb,Almheiri:2019yqk} then provides a method of deriving the entanglement entropy of a region, R, in the non-gravitational system. The island rule is that this entropy is obtained by extremizing,
\be
\label{islandrule}
S(R) = \text{min}\left[\text{ext}\left({A(\partial I) \over 4 G} + S_{\text{bulk}}(I \cup R) \right) \right],
\ee
where $I$ is a part of the gravitating system. The natural extension of the subregion duality proposal suggests that operators in $R$ can describe the physics of $I$. 

The formula \eqref{islandrule} has been carefully derived in JT gravity using a replica trick \cite{Almheiri:2019qdq}. In other settings, the formula has been justified but again when $R$ is in the non-gravitational region \cite{Bousso:2020kmy}. In some parts of the literature \eqref{islandrule} is directly applied even when $R$ is in a region with dynamical gravity. It has already been pointed out in \cite{Geng:2020fxl,Laddha:2020kvp} that since the entanglement entropy is a fine-grained quantity, even the presence of weak gravity can alter its magnitude. Therefore \eqref{islandrule} is not directly applicable to settings where gravity is dynamical everywhere. The puzzle that we describe below adds additional evidence for this claim.

The puzzle arises from the following simple observation.
\begin{observation}
Islands are the only known example of entanglement wedges that do not extend to the asymptotic boundary of the gravitating spacetime. They are disconnected from the region $R$ in the sense that not even a spatial geodesic from $I$ can reach $R$ without passing through the complement $\overline{I}$ of the island.
\end{observation}

Here we are using $\overline{I}$ to denote the complement of the island in the gravitating system to be consistent with the notation used above.

In light of the discussion of section \ref{secasymptotic}, this leads to a puzzle because in an ordinary theory of gravity, there is no way to dress operators in $I$ without making reference to operators in $\overline{I}$. This is an obstacle to making operators in $I$ invariant under diffeomorphisms. This puzzle can be made sharp as follows.

Let $\phi(P)$ be a Hermitian scalar field operator that probes physics at a point $P \in I$. Consider the unitary operator $U = e^{i \lambda \phi(P)}$ where $\lambda$ is a small parameter introduced for convenience below. Then since $\phi(P)$ is described by an operator in $R$, 
the unitary operator $U$ should leave the expectation value of all operators in the algebra $\overline{\mathcal{A}(R)}$ unchanged. This can be seen as follows. Let $|\Psi \rangle$ denote the state of the entire system, including the bath. Let $A_{\bar{R}} \in \overline{\mathcal{A}(R)}$. We then expect to have,
\be
\label{unchangedabar}
\langle \Psi| U^{\dagger} A_{\bar{R}} U |\Psi \rangle = \langle \Psi | A_{\bar{R}} U^{\dagger} U |\Psi \rangle = \langle \Psi| A_{\bar{R}} |\Psi \rangle,
\ee
where we have used the commutator \eqref{commutcomplement} and the unitarity of $U$.
Since operators in $\overline{I}$ are dual to operators in $\overline{R}$, this means that the action of $U$ should also leave the expectation value of all operators in $\overline{I}$ unchanged.

But, in an ordinary theory of gravity, the Gauss law tells us that the asymptotic metric near the boundary of AdS (which is in $\overline{I}$) measures the energy of the bulk, which includes $I$. 
Let $A_{\bar{I}}$ be another simple operator that acts on the complement of the island. Then, using equation \eqref{extrapolatedict} and the Gauss law we find that,
\be
\label{heisenbergeom}
\begin{split}
\lim_{r \rightarrow \infty} r^{\dt-2} \int_{(\partial\text{AdS})} {\partial \over \partial \lambda} \left. \langle \Psi | U A_{\bar{I}} h_{00} U^{\dagger} | \Psi \rangle \right|_{\lambda = 0} &= \lim_{r \rightarrow \infty} r^{\dt-2} \int_{\partial(\text{AdS})} {\partial \over \partial \lambda} \left.\langle \Psi | A_{\bar{I}} U h_{00} U^{\dagger} | \Psi \rangle \right|_{\lambda=0}\\
&= \lim_{r \rightarrow \infty} r^{\dt-2} i \int_{\partial(\text{AdS})} \langle \Psi | A_{\bar{I}} [\phi(P), h_{00}] | \Psi \rangle \\
&= {-16 \pi G \over \dt} \langle \Psi | A_{\bar{I}} {\partial \phi(P) \over \partial t} |\Psi \rangle,
\end{split}
\ee
which is different from \eqref{unchangedabar}.

Physically, equation \eqref{heisenbergeom} has a simple interpretation. The unitary $U$ inserts a small excitation in the region $I$. The metric at infinity should be able to measure the energy of this excitation. Note that \eqref{heisenbergeom} involves an insertion of the gravitational constant and so this commutator appears at leading nontrivial order in perturbation theory. 

Observe that if one takes the nongravitational limit, the right hand side of \eqref{heisenbergeom} vanishes. This is why it is possible to discuss local measurements when gravity can be neglected. 

\paragraph{\bf Black holes.}
We would like to make a few important comments about islands in the presence of black holes. First note that \eqref{heisenbergeom} is not just about the {\em expectation value} of the energy. When there is a black hole in the bulk, one might attempt to consider operators that somehow ``extract'' energy from inside the black hole and ``insert'' it in some part of the island that is outside the black hole.
However, even such an operator would have to change the distribution of energy in the island, so it would not commute with the metric near infinity. This nonzero commutator can be detected by the insertion of an appropriate operator $A_{\bar{I}}$ in the correlator \eqref{heisenbergeom}.

\begin{figure}
\begin{center}
\includegraphics[height=0.3\textheight]{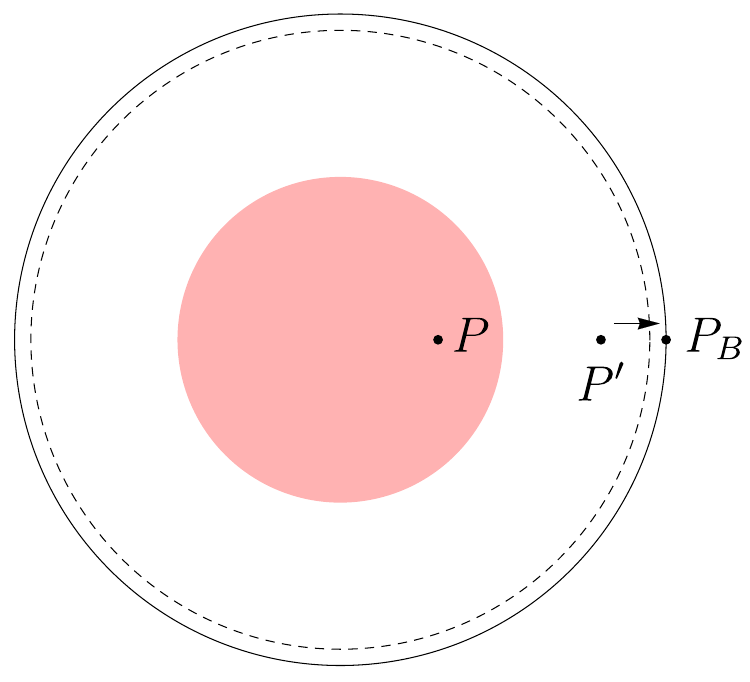}
\caption{\em An excitation at the point $P$ inside the island (pink shaded disk) can be detected using a two-point correlator outside the island. The two-point correlator involves an integral of the asymptotic metric (indicated by the dashed line) and another operator obtained by taking the limit of point $P'$ to $\pbound$ on the boundary of the gravitational region. The bath is not shown in this figure, which shows a time slice of AdS. \label{figtwopt}}
\end{center}
\end{figure}

Let us consider a more explicit example. Let the state $|\Psi \rangle$ correspond to a black hole that has equilibriated with a bath at the same temperature. Assume that the field $\phi$ describes a scalar excitation of mass $\mu$ and let $\Delta = {{\dt \over 2}} + \sqrt{\mu^2 + {\dt^2\over 4}}$. Then an explicit choice of $A_{\bar{I}}$ that leads to a nonzero value for the correlator in equation \eqref{heisenbergeom} is simply,
\be
\label{achoice}
A_{\bar{I}} = \lim_{P' \rightarrow \pbound} r^{\Delta} {\partial \phi(P') \over \partial t},
\ee
where $P'$ is a point in $\overline{I}$ with radial coordinate $r$ that is taken to a point on the boundary of the gravitational region $\pbound$ and scaled up to yield a finite operator. (See Figure \ref{figtwopt}.) 
In the absence of the unitary operator in equation \eqref{heisenbergeom} we find that,
\be
\lim_{r \rightarrow \infty} r^{\dt-2} \int_{\partial\text{AdS}} \langle \Psi |A_{\bar{I}} h_{00} | \Psi \rangle = 0.
\ee

This equation follows because in an equilibrium state, there is no preferred direction of time and so correlators of the Hamiltonian and a time-derivative of a bulk scalar field vanish; even the fluctuations of such a correlator are exponentially small. On the other hand, in the presence of the unitary operator we see that the final result in equation \eqref{heisenbergeom} becomes a two-point {\em Wightman function} of the time-derivative of the field---with one insertion inside the island and another insertion near the boundary. This Wightman function does not vanish even when one point is inside the horizon and the other is outside the horizon.

We emphasize that our objective here is not to completely reconstruct the state $|\Psi \rangle$, which would require exponential precision. It is only to show that {\em some} simple excitations, including those produced by unitary operators involving bulk fields, can be detected from outside the island. These simple excitations should be in the ``little Hilbert space'' or ``code subspace'' built about the state $|\Psi \rangle$.

Moreover, since we are considering only simple excitations it is safe to evaluate the correlator in \eqref{heisenbergeom} in the black-hole background. The simple correlators that appear in \eqref{heisenbergeom} do not receive significant contributions from nonperturbatively suppressed branches of the wavefunction, which might be important for the computation of very high-point correlators. In particular, the perturbative calculation leading to \eqref{heisenbergeom} cannot be invalidated by considering exotic configurations where $\phi(P)$ can be dressed to the non-gravitational bath using wormholes that bypass the complement of the island.\footnote{In the next section, we will see that when the island and the bath are realized in a higher-dimensional doubly-holographic setting, $\phi(P)$ can be dressed to the non-gravitational bath through the higher dimension. But, in this setting, the AdS$_{\dt+1}$ theory of gravity is massive.}

All of the comments above hold if $|\Psi \rangle$ corresponds to a state of an eternal black hole coupled to a bath at the same temperature. The only subtlety is that since the gravitational part of the geometry has two asymptotic regions the boundary of AdS in \eqref{heisenbergeom} should be interpreted as the union of the two asymptotic boundaries.

One might wonder if the integrated asymptotic metric is the only operator that causes a problem in interpreting the island as an entanglement wedge. This is not the case. Even if one attempts to ``discard'' the asymptotic metric from the algebra of operators in $\overline{I}$ at leading order, it will reappear in the algebra at subleading order \cite{ElShowk:2011ag}. This is simply due to the fact that the OPE of boundary operators produces the boundary stress tensor. In the bulk, this means that the algebra of other asymptotic operators produces the asymptotic metric. So, for consistency with \eqref{heisenbergeom}, the operator $U$ must fail to commute with other operators in $\overline{I}$ although such commutators may appear only at subleading orders in perturbation theory.

For the reader who would like additional details, we note that the detection of simple excitations about black holes has been studied previously in the literature. In \cite{Papadodimas:2013jku} (see section 5) a class of excitations of black holes leading to ``near-equilibrium'' states was studied and it was shown how they could be detected. Excitations of the island in the region outside the black hole horizon correspond to the ``near-equilibrium'' states of \cite{Papadodimas:2013jku}. In \cite{Papadodimas:2015jra} (see section 8.1) and also \cite{Papadodimas:2017qit}, a more general class of excitations of the black-hole interior was studied and it was shown how correlators of the Hamiltonian and other operators could be used to detect them as well.

We have therefore arrived at the following puzzle.
\begin{puzzle}
The Gauss law suggests that the action of any operator in the island must be accompanied by a disturbance in the metric outside the island. This is in contradiction with the idea that operators in the island are described by operators in $R$ and commute with operators in the complement of the island that are described by operators in $\overline{R}$.
\end{puzzle}
We remind the reader that we use the phrases ``island'' and ``Gauss law'' in the precise sense described in section \ref{definition}. 

Note that our puzzle pertains to whether the islands can constitute consistent entanglement wedges. If the island rule \eqref{islandrule} is used merely as a trick to compute the entropy without a concomitant claim that the island is the entanglement wedge of the radiation region, then our puzzle would not apply. However, we do not consider this possibility further since the proof of subregion duality follows directly from the entropy formula \cite{Jafferis:2015del,Dong:2016eik}, and so the computation of the entropy and the determination of the entanglement wedge cannot usually be separated. 

We also note that the mere existence of solutions to \eqref{islandrule} in standard theories of long-range gravity \cite{Ghosh:2021axl,Krishnan:2020fer,Hollowood:2021nlo,Wang:2021mqq,Wang:2021woy} cannot be used to conclude that such theories must exhibit a Page curve or have islands as entanglement wedges. The physical interpretation of such solutions is unclear since the island rule has not been justified in standard theories of gravity. In particular, the Gauss-law puzzle above implies that even if an island is obtained as a geometric solution to a minimization problem in a theory of long-range gravity, it does not constitute a consistent entanglement wedge.

We have formulated a puzzle above using the gravitational Gauss law. A similar puzzle can be formulated using the Gauss law in gauge theories. The action of a charged operator in the island must be accompanied by a change in the gauge field outside the island. The puzzle is gauge theories is less acute than it is in gravity since gauge theories contain local gauge-invariant operators and there is no obstacle to localizing such operators in the island. In standard theories of gravity, as we have already explained, there are no local gauge-invariant operators.

\section{A resolution using massive gravity \label{secmassiveresolve}}
In this section we describe how the puzzle of section \ref{secpuzzle} can be resolved in the setting of massive gravity. We will argue that the puzzle of section \ref{secpuzzle} does not appear in this scenario. 

\begin{figure}[!ht]
\begin{center}
\includegraphics[height=0.3\textheight]{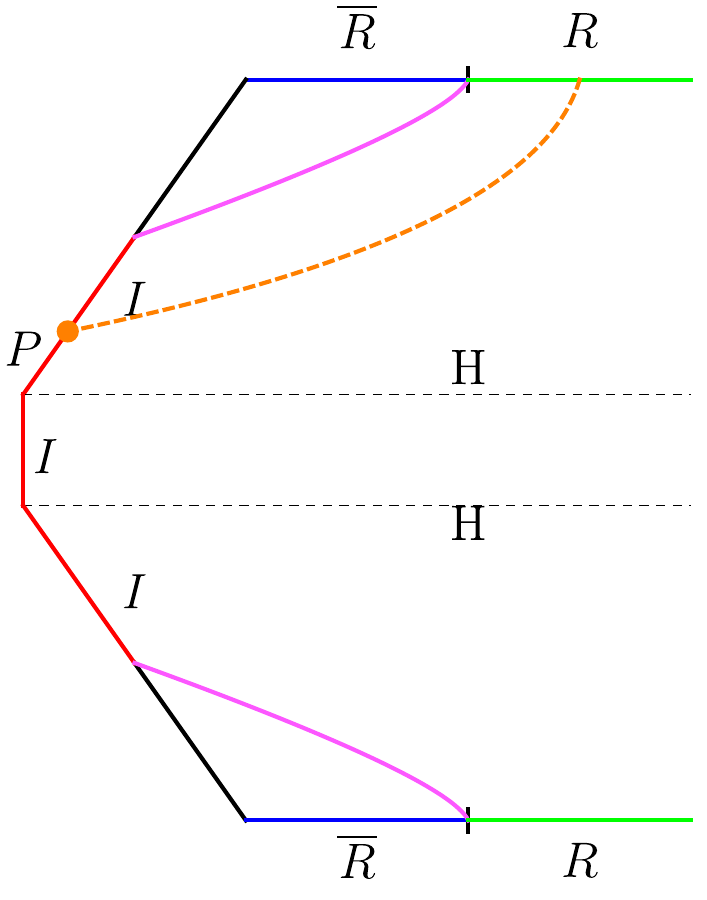}
\caption{\em In the higher-dimensional setup, an operator at point $P$ can be dressed to the boundary through the higher-dimension. The setup is the same as that of Figure \ref{fighigherd} and this Figure illustrates how an operator at $P$ can be dressed to the boundary while bypassing the entanglement wedge of $\overline{R}$. In this figure both the horizontal and vertical directions are spatial. \label{fighigherdwithblob}}
\end{center}
\end{figure}

Because there are few well-understood examples of massive gravity and because it has naturally occurred in the context of entropy calculations in higher dimensions, our starting point is the Karch-Randall setup that has been used to study islands in higher dimensions that we reviewed earlier. Here one embeds a $d$-dimensional AdS brane in a $(d+1)$-dimensional AdS bulk. The boundary dual to this geometry is believed to be a BCFT$_{d}$, a CFT$_{d}$ on a space with a boundary and with conformal boundary conditions for CFT fields as one approaches this boundary \cite{Takayanagi:2011zk}. In addition, the boundary of this half-space can support additional degrees of freedom and is sometimes referred to as a ``defect'' \cite{Gaiotto:2008sa}. One can consider a thermofield double state of two such BCFTs which is then dual to an eternal black hole with two asymptotic boundaries and a brane that runs from one boundary to the other. (See Figure \ref{fighigherdwithblob}.)

It is believed that the correct bulk generalization \cite{Almheiri:2019yqk} of the holographic entanglement entropy prescription is obtained through the following generalization of the homology constraint: the entropy of a region $R$ on the boundary is given by formula \eqref{holprescription}, where one is allowed to consider all those surfaces $X$ so that $E$ has no boundaries except for $R$, $X$, {\em and a possible portion of the brane}. In particular, if one takes $R$ to be the union of a region on one asymptotic boundary with a similar region on the other asymptotic boundary, then at late enough times, one finds a phase transition for the surface $X$ between what is called a Hartman-Maldacena surface \cite{Hartman:2013qma}, which runs from $\partial R$ on one boundary to $\partial R$ on the other boundary, and a second surface that runs from $\partial R$ to the brane as shown in Figure \ref{fighigherdwithblob}.\footnote{There are technically two copies of this second surface, with each residing in a respective exterior patch of the black hole.}

For this entanglement wedge, it is clear that the puzzle of section \ref{secpuzzle} does not arise. If we consider an operator that acts near the brane as shown in Figure \ref{fighigherdwithblob}, then this operator can be dressed to the asymptotic boundary in region $R$ entirely within the entanglement wedge $E(R)$ without ever entering its complement. We can ensure that the operator commutes with all operators on $\overline{R}$ but not that it commutes with operators in $R$. On the other hand, as Figure \ref{fighigherdwithblob} shows, the entanglement wedge does not contain an ``island'' (in the sense of subsection \ref{definition}). This entanglement wedge is just a conventional entanglement wedge of the kind described in section \ref{secasymptotic} comprising only regions that extend to the asymptotic boundary. So there is no puzzle involving the Gauss law, just as there is no puzzle with conventional entanglement wedges in AdS/CFT.

\begin{figure}[!ht]
\begin{center}
\includegraphics[height=0.3\textheight]{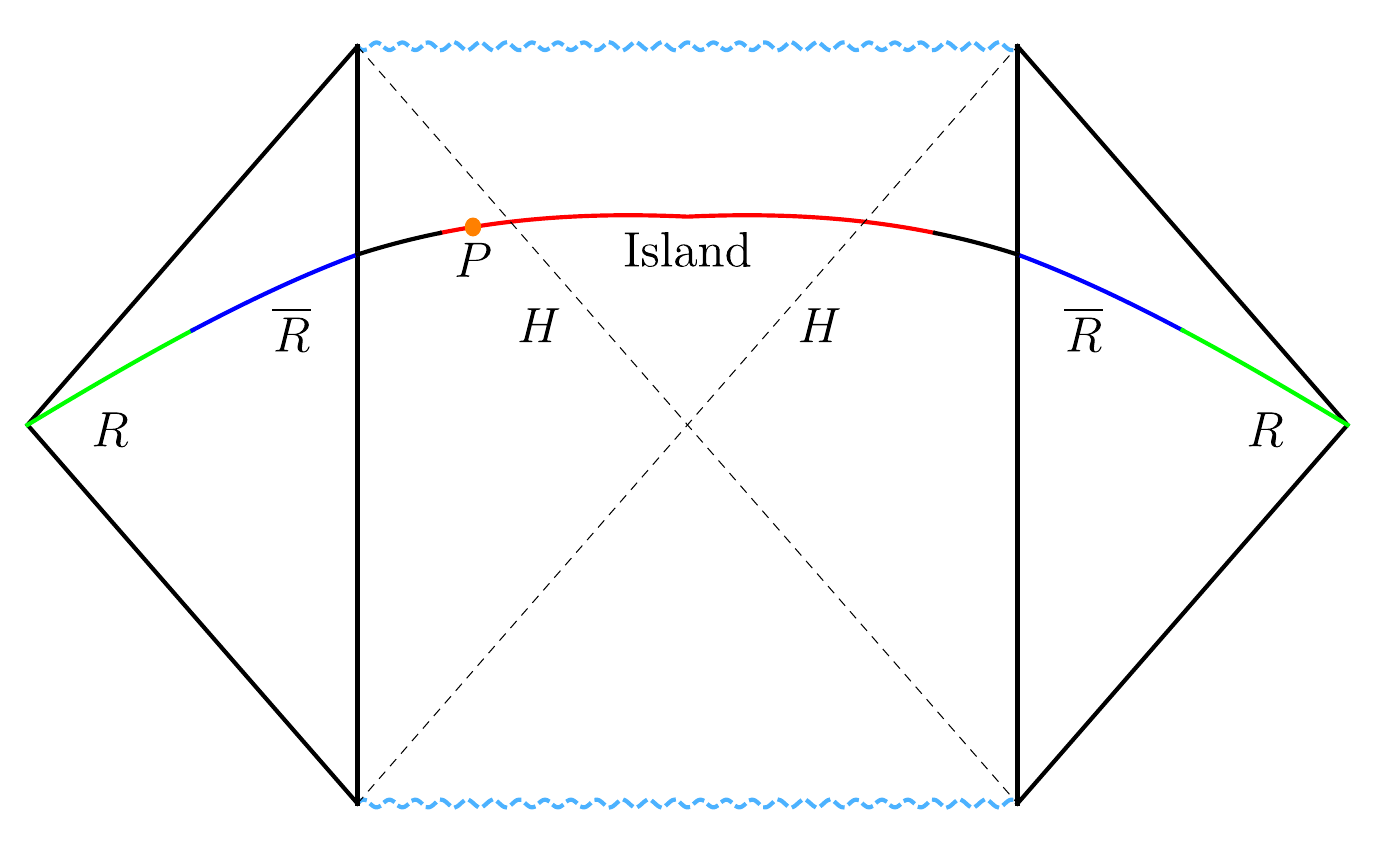}
\caption{\em A lower-dimensional picture of the setup of Figure \ref{fighigherdwithblob}. This is a spacetime diagram like Figure \ref{figlowerd}.  In the lower-dimensional description, an operator at point $P$ in the island cannot be dressed to the boundary without affecting the region outside the island. So the island cannot constitute a consistent entanglement wedge in a theory where the Gauss law applies. The lower-dimensional description of Figure \ref{fighigherdwithblob} involves massive gravity where the Gauss law does not hold. In this figure, the horizontal direction is spatial and time runs along the vertical direction. \label{islandlower}}
\end{center}
\end{figure}

An apparent puzzle appears because the configuration under discussion admits yet another description, obtained by dualizing the gravitational theory on AdS$_{d+1}$ to a gravitational theory in AdS$_d$ coupled to a non-gravitational bath with transparent boundary conditions. In this description, the entanglement wedge that we have discussed above is shown in Figure \ref{islandlower}. The origin of the term ``island'' is now clear since the part of the entanglement wedge that terminated on the brane in the higher-dimensional description of Figure \ref{fighigherdwithblob} now appears be disconnected from the asymptotic region in this lower-dimensional picture. But if one considers the action of a unitary operator at the point $P$ in the lower-dimensional picture, one might wonder how this avoids the puzzle of section \ref{secpuzzle}.

Since the higher-dimensional picture does not involve any violation of the Gauss law, the lower-dimensional picture must also be consistent. A resolution to the apparent Gauss-law puzzle {\em must} therefore lie in the details of the dimensional reduction. 

A notable aspect of the dimensionally-reduced picture is that the lower-dimensional theory of gravity is always {\em massive}. 
This can be understood from the perspective of the gravitational theory on the brane \cite{Porrati:2003sa,Porrati:2001gx,Porrati:2002dt}.

But another simple way to understand the origin of the mass is from the boundary. In the absence of the bath, the theory of gravity on the AdS brane is dual to a conformal field theory with a conserved stress tensor. The coupling to the bath leads to the nonconservation of the stress tensor on the boundary. This allows the stress tensor to pick up an anomalous dimension \cite{Aharony:2006hz} which, in the bulk, corresponds to a massive graviton. This interpretation is important because it would generalize to any such theory coupled to a bath such as the one analyzed in \cite{Penington:2019npb}. The coupling should always generate a mass for a propagating graviton. We will show that this mass will always be present in consistent scenarios that include an island. Note that even if, as is sometimes suggested, the coupling to the bath is turned off after evaporation, islands always appear concomitantly with a mass for the graviton.

 As we further elaborate below, the mass of the graviton resolves the apparent Gauss-law puzzle for the simple reason that the constraints of massive gravity cannot be integrated to obtain a Gauss law of the form that exists in the massless theory. Consequently, in a theory of massive gravity, even when a region is surrounded by its complement, it {\em is possible} to modify the state of the region without modifying the state of its complement. This is an example of how massive gravity can have {\em qualitatively different} properties from massless gravity.

In flat space, the gravitational force law changes discontinuously as the mass of the graviton goes to zero, and this is known as the vDVZ discontinuity \cite{vanDam:1970vg,Zakharov:1970cc}. In AdS, there is no such discontinuity in the gravitational force law. However, there is still a qualitative difference in the way massless and massive theories of gravity {\em store quantum information.} It is this difference that allows islands to exist in theories of massive gravity but not in theories of massless gravity where the Gauss law applies.

The rest of this section is divided into two parts. First, we provide a simple explanation of our main idea in subsection \ref{secflatlinear}, which illustrates the difference between massive and massless gravity in flat space. This discussion avoids technical details but is sufficient to understand the point which we wish to emphasize.

Then, in subsection \ref{subsecconstraintbrane}, we explore the form of the gravitational constraints in the higher-dimensional AdS$_{d+1}$ theory, focusing on the so-called Hamiltonian constraint. We linearize the Hamiltonian constraint and perform a Kaluza-Klein reduction of the constraint in a braneworld geometry. We show that these constraints of the $(d+1)$-dimensional bulk theory go over to the constraints of $d$-dimensional massive gravity in subsection \ref{subsecdimred}. We show in subsection \ref{subsecdimred} that the mass spectrum obtained in this manner is precisely the known mass spectrum of Karch-Randall braneworlds.

\subsection{Massless vs. massive constraints in flat-space linearized gravity \label{secflatlinear}}

In this section, we illustrate the difference between massless and massive gravity in a simple setting: linearized gravity in flat space. 

Consider a theory of gravity coupled to matter in flat space. We are interested in a state that is close to the vacuum with some matter energy density $\rho$. Since we would like to examine the constraints of the theory, we focus on a {\em spatial slice} at a specific instant of time. We take the spatial components of the metric on this slice to be $\delta_{i j} + h_{i j}$. Note that $i,j$ do not run over the time coordinate. Then this metric fluctuation and the energy density are not entirely uncorrelated. In standard massless gravity, at the linearized level, they must obey the constraint,
\be
\label{localcurrent}
-\partial_j \partial_j h_{i i} + \partial_j \partial_i h_{i j} = {16 \pi G} \rho,
\ee

This constraint follows from the linearized $TT$ component of the Einstein equation, but it can also be derived from a standard canonical analysis in the Hamiltonian formalism. 

Integrating this constraint over a volume $V$ leads to,
\be
\label{gausslaw}
{1 \over 16 \pi G} \int_B d^{d-1} x\ n_j(\partial_i h_{i j} - \partial_j h_{i i} ) = \int_{V} d^d x\ \rho. 
\ee

Here the left integral is performed over the boundary $B$ of the region $V$, and $n_j$ is the unit normal vector of the boundary. On the right hand side, we have a bulk integral over the entire region $V$. 
This is just the standard Gauss law, used in the sense of section \ref{definition}, which relates the total energy to the asymptotic metric.

We note an aspect of equation \eqref{localcurrent} that will be relevant below. It is convenient to decompose the metric perturbation, following ADM \cite{Arnowitt:1962hi,Kuchar:1970mu}, into a ``longitudinal'' (L) component, a ``transverse traceless'' (TT) component and what we call a ``T'' component, 
\be
\label{admdecomposition}
h_{i j} = \hlo_{i j} + \htt_{i j} + \htr_{i j},
\ee
where,
\be
\hlo_{i j} = \partial_{(i} \epsilon_{j)},
\ee
for some vector field $\epsilon_j$ and,
\be
\partial_i \htr_{i j} = \partial_{i} \htt_{i j} = 0; \qquad \htt_{i i} = 0.
\ee

For an explicit decomposition of any metric perturbation into \eqref{admdecomposition} see the nice discussion in \cite{kuchar1991problem}.

Both $\hlo_{i j}$---which corresponds to spatial diffeomorphisms of the slice---and $\htt_{i j}$---which parameterizes the dynamical graviton---drop out of equation \eqref{localcurrent}. Thus, equation \eqref{localcurrent} constrains only the T component of the metric perturbation. We will return to the relevance of this observation when we study the constraints in massive gravity.

We would like to make a few comments.
\begin{enumerate}
\item
The equation above was derived in the linearized approximation. However, when $V$ is taken to be an entire Cauchy slice, the term on the left hand side of equation \eqref{gausslaw} turns into the famous ADM Hamiltonian \cite{Arnowitt:1962hi}. So when $B$ is the large-$r$ region of the Cauchy slice, the left hand side of equation \eqref{gausslaw} is the {\em definition} of the energy, even in the full theory of general relativity.
\item
Upon the insertion of an excitation in the bulk, which changes the integral of $\rho$, the constraint {\em forces} a concomitant change in the metric at infinity. So even if the excitation appears to insert energy only in a bounded region, the metric around that region all the way up to infinity must be changed.
\item
In the classical theory, it is possible to ``move'' energy from one spot to another in the bulk while keeping the metric unchanged outside a large ball. This is guaranteed in the classical theory by the Birkhoff theorem and its generalization by Corvino and Schoen \cite{Corvino:2003sp}. Such a possibility does {\em not} exist in the quantum theory. In the quantum theory, the boundary Hamiltonian not only knows about the expectation value of the energy but also about {\em moments} of the energy and of {\em correlators} of the energy with other observables. This information is enough to ensure that it is impossible to change even the distribution of the energy in the bulk region without affecting at least some correlator of the boundary metric with other boundary degrees of freedom. This was termed the principle of ``holography of information'' in \cite{Raju:2020smc}.
\item
The constraints of the theory hold on a single Cauchy slice, and so they hold even if the spacetime has a horizon, since the horizon arises from the global causal properties of the spacetime. In particular, the constraints are important even in the presence of black holes. The insertion of an excitation in the interior of the black hole still changes the asymptotic Hamiltonian. As mentioned above, it can be shown that simple excitations in the interior can be detected by correlators of the asymptotic Hamiltonian and other operators as described in \cite{Papadodimas:2017qit}. (See also section 8 of \cite{Papadodimas:2015jra}.)
\end{enumerate}

We now turn to massive gravity. Even in massive gravity, we find that if one studies a state close to the Minkowski vacuum then the matter energy-density and the metric perturbation are related by the following constraint \cite{Hinterbichler:2011tt}:
\be
\label{massiveconstraint}
-\partial_j \partial_j h_{i i} + \partial_j \partial_i h_{i j} + m^2 h_{i i}= 16 \pi G \rho.
\ee

We note an important difference between equation \eqref{massiveconstraint} and \eqref{localcurrent}. The left hand side of \eqref{massiveconstraint} is {\em not} a gradient. Consequently the integral of the energy-density over a volume cannot be expressed in terms of the integral of the boundary metric and its derivatives. This is why the energy of the state is {\em not} just given by a boundary term and there is no analogue of the Gauss law (as defined in subsection \ref{definition}) in theories of massive gravity. 

We emphasize that equation \eqref{massiveconstraint} should not be thought of simply as a ``screened'' version of equation \eqref{localcurrent}. This is because equation \eqref{massiveconstraint} now involves $\hlo_{i j}$ from the decomposition \eqref{admdecomposition} and is not just a constraint on $\htr_{i j}$. The constraint relates the longitudinal mode, which is now an additional degree of freedom, to the other modes of the graviton. 

Therefore, at least at this linearized level and when the graviton has a mass, it {\em is} possible to insert energy at a location while keeping the metric far away unchanged. We simply use the $m^2 h_{i i}$ term to compensate for the change in $\rho$. 
So, it is possible for energy to ``appear'' in the middle of a bounded region, i.e. for $\rho$ to change, without an alteration of the metric at the boundary of that region. Some more discussion of this effect from a phenomenological perspective can be found in \cite{Goldhaber:2008xy}.

We are not aware of any detailed analysis of quantum wavefunctionals that satisfy the constraint \eqref{massiveconstraint} or any analysis that carefully accounts for the effects of possible nonlinearities. However, since the Hamiltonian is not a boundary term in massive gravity we do not see any \textit{a priori} obstruction to preparing ``split states'' in massive gravity: states that differ within a bounded region but are identical outside that region.
 
This suggests that the puzzle of section \ref{secpuzzle} is removed when the bulk graviton has a mass. The island is surrounded by its complement. But if the bulk theory of gravity is massive, it is possible for degrees of freedom in the island to correspond to degrees of freedom in a disconnected ``radiation'' region. An insertion of energy, performed via the action of a unitary operator in the radiation region, does not need to modify the metric in the complement of the island.

We now turn to a more detailed investigation of the constraints when the massive graviton is realized through dimensional reduction of a higher-dimensional gravitational theory.

\subsection{Gravitational constraints for AdS spacetimes with branes \label{subsecconstraintbrane}}
The diffeomorphism-invariance of gravity leads to a set of constraints that must be obeyed by valid wavefunctionals even when we go beyond the linearized approximation. These constraints are commonly divided into what are called the ``momentum constraints'' and the ``Hamiltonian constraint'' \cite{DeWitt:1967yk}. We review these constraints below in the context of asymptotically AdS spacetimes that support a brane, and we explain how they directly lead to a Gauss law in the higher-dimensional spacetime. We then dimensionally reduce these constraints on the brane and show how the mass of the graviton appears in the dimensionally-reduced constraint. This supports the idea introduced above that the mass is key to the consistency of islands on the brane. The analysis in this section has some overlap with the analysis of \cite{Chowdhury:2021nxw} where the reader will find further details.

The analysis of gravitational constraints is commonly performed after a $d+1$ split of the metric, so that the line element takes the form,
\be
ds^2 = -N^2 dt^2 + \gamma_{i j} (d x^i + N^i d t) (d x^j + N^j d t).
\ee

Here, $N$ is called the ``lapse function,'' $N^i$ is called the ``shift vector,'' and $\gamma_{i j}$ is the metric on spatial $d$-dimensional slices. In addition, the theory might contain matter fields that we simply denote by $\phi_{\text{matter}}$ since they will play a limited role in the analysis.

The nontrivial constraint equation in gravity is the so-called Hamiltonian constraint which tells us that any valid state of the theory must satisfy 
\be
\label{hamiltconst}
{\cal H} = {1 \over 2 \sqrt{\gamma}} \left(\gamma_{i k} \gamma_{j l} + \gamma_{i l} \gamma_{j k} - {2 \over d-1} \gamma_{i j} \gamma_{k l} \right) \Pi^{i j} \Pi^{k l} - \, \sqrt{\gamma} R + {2 \sqrt{\gamma}} \Lambda + 16 \pi G {\cal H}_{\text{matter}} = 0.
\ee
Here ${\cal H}_{\text{matter}}$ is the Hamiltonian density of the matter sector. $R$ is the $d$-dimensional Ricci scalar and we have included a possible cosmological constant $\Lambda$. $\Pi^{i j}$ is the momentum conjugate to the metric. In the quantum theory, $\Pi^{i j}$ is represented as $-i{\partial \over \partial \gamma_{i j}}$, but in the classical limit
the conjugate momentum is related to the extrinsic curvature of the spatial $d$-dimensional slices $K_{i j}$ via,
\be
\Pi^{i j} = - \sqrt{\gamma} \left(K^{i j} - \gamma^{i j} \gamma_{k l} K^{k l} \right).
\ee

The constraint \eqref{hamiltconst} must be obeyed by any valid state and in a standard theory of long-range gravity it implies a Gauss law. To illustrate the physics, we show how this Gauss law emerges when the constraint is linearized for a family of simple states.  
We assume that we are considering a vacuum background solution to \eqref{hamiltconst} with,
\be
\label{backgroundsol}
\gamma_{i j} = g_{i j}; \qquad \Pi_{i j} = 0; \qquad {\cal H}_{\text{matter}} = 0; \qquad N^i = 0. \qquad \text{[Background~solution]}
\ee
We will use $N$ to denote the value of the lapse in this background. Although equation \eqref{backgroundsol} describes a simple background, the final expression we will obtain for the energy, as measured from the boundary, will be more general.

We are interested in a nearby solution with a nonzero matter energy-density $\delta \hmat$. To see how the metric must respond to this energy density, we expand the metric about this background as,
\be
\label{hdefn}
\gamma_{i j} = g_{i j} + h_{i j},
\ee
where $g_{i j}$ is the background metric and $h_{i j}$ is the metric fluctuation. After some algebra we find that the linear term in $h_{i j}$ is of the form
\be
\label{linearization}
\sqrt{\gamma} (R - 2 \Lambda) = \sqrt{g} \left[ -h^{i j} R^{(g)}_{i j} + {1 \over 2} \ h^{i}_{i} R^{(g)} - h^{i}_{i} \Lambda + \nabla_{i} \nabla_{j} h^{i j} - \nabla^{j} \nabla_{j} h^{i}_{i} \right] + \ldots
\ee
where $\ldots$ indicates higher-order terms, $\nabla_{i}$ is the covariant derivative with respect to $g_{i j}$, $R^{(g)}_{i j}$ and $R^{(g)}$ are the background Ricci tensor and scalar, and all indices are raised using $g^{i j}$.

This may not immediately seem like a total derivative. However, for the background solution \eqref{backgroundsol}, we have the identity,
\be
G^{(g)}_{i j} = R_{i j}^{(g)} - {1 \over 2} g_{i j} R^{(g)} - {1 \over N} \left(\nabla_i \nabla_j - g_{i j} \nabla^2 \right) N.
\ee

The vacuum Einstein equations are,
\be
G^{(g)}_{i j} + \Lambda g_{i j} = 0,
\ee
so therefore we can write \eqref{linearization} as,
\be
\begin{split}
\sqrt{\gamma} (R - 2 \Lambda) &= {-\sqrt{g} \over N} \left(\nabla^i \nabla^j N - g^{i j} \nabla^2 N \right) h_{i j} + \sqrt{g} \left(\nabla_{i} \nabla_{j} h^{i j} - \nabla^{j} \nabla_{j} h^{i}_{i} \right) \\
&= {\sqrt{g} \over N} \left[\nabla_i \left(N \nabla_j h^{i j} \right) - \nabla_i \left( h^{i j} \nabla_j N \right) - \nabla_i \left(N \nabla^i h^{j}_{j} \right) + \nabla_i \left(h^{j}_{j} \nabla^i N \right)\right].
\end{split}
\ee

In the last line above we have also relabeled some dummy indices. The constraint can now be written as,
\be
\label{constraintcurved}
{1 \over 16 \pi G} \sqrt{g} \nabla_i J^i = N \sqrt{g} \delta \hmat,
\ee

Here we have defined
\be
\label{conservcurrentdef}
J^i \equiv N \nabla_j h^{i j} - h^{i j} \nabla_j N - N \nabla^i h^{j}_{j} + h^{j}_{j} \nabla^i N .
\ee

We now integrate both sides of \eqref{constraintcurved} on the entire Cauchy slice. We denote the boundary of the slice by $S_{\infty}$ and the unit normal vector to the boundary by $n_i$. The left hand side of \eqref{constraintcurved} then integrates to a quantity that we denote by $E$,
\be
\label{bdryintegral}
E \equiv {1 \over 16 \pi G} \int_{S_{\infty}}d^{d-1} x\,\sqrt{g}\,n_i J^i,
\ee
and \eqref{constraintcurved} reduces to,
\be
\label{intconstraintcurved}
E = \int d^d x\, N\sqrt{g}\delta \hmat.
\ee

This is the generalization of the Gauss law \eqref{gausslaw} to curved space, which tells us that an insertion of energy density at a point in the bulk {\em must} be accompanied by a change of the boundary metric in order to satisfy the constraints.

The expression above was derived in the presence of some approximations. In the full theory, we need to account for higher-order terms in the expression \eqref{hamiltconst} and also account for the energy of transverse-traceless gravitons themselves. Nevertheless, even in the full theory of general relativity, the boundary integral \eqref{bdryintegral} provides the correct definition of the energy. It can be seen that this coincides with the expression of \cite{Hawking:1995fd} and \cite{Balasubramanian:1999re} for energy in asymptotically anti-de Sitter spacetimes and can also be shown to be equivalent to the expression that follows \eqref{extrapolatedict}. We refer the reader to \cite{Chowdhury:2021nxw} for details.

Equation \eqref{bdryintegral} also provides the correct expression for the energy in the quantum theory. In the quantum theory, denoting the matter fields collectively by $\phi_{\text{matter}}$, states are represented by wavefunctionals $\Psi[\gamma_{i j}, \phi_{\text{matter}}]$. Valid states must satisfy \eqref{hamiltconst} in the sense that,
\be
{\cal H} \Psi[\gamma_{i j}, \phi_{\text{matter}}] = 0.
\ee
where $\Pi^{i j} = -i {\partial \over \partial \gamma_{i j}}$. The value of the energy for any valid state is then again given by the expression \eqref{bdryintegral}.

\subsection{Dimensional reduction of constraints in warped geometries \label{subsecdimred}}

We now specialize the constraints above to the geometry with branes. We would like to show that the local constraint \eqref{constraintcurved} goes over, after dimensional reduction, to the local constraint of a theory of massive gravity. The mass spectrum of the lower-dimensional graviton has been studied in geometries without black holes and so, for simplicity, we consider such geometries here. The results that we derive below will have a clear generalization.

We consider a family of ``warped geometries'' in $d+1$ dimensions that can be written in the form,
\be
ds^2 = d\wr^2 + e^{2 A(\wr)} \left(-\bar{N}^2 d t^2 + \bar{\gamma}_{\bar{i} \bar{j}} d x^{\bar{i}} d x^{\bar{j}} \right).
\ee

\begin{figure}
\begin{center}
\includegraphics[height=0.3\textheight]{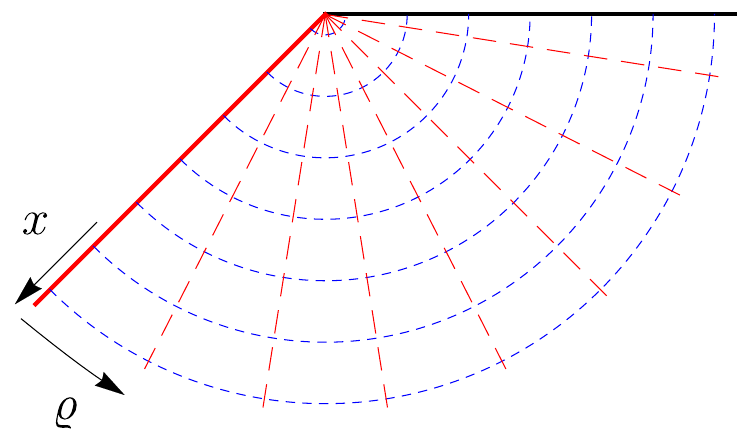}
\caption{\em An AdS$_d$ brane (thick red line) in AdS$_{d+1}$. The conformal boundary is the thick black line. Lines of constant $\wr$ are red dashed lines and coordinates transverse to $\wr$ are constant along the blue dashed circles.\label{figdimred}}
\end{center}
\end{figure}

Our notation for the coordinates is as follows. We consider a brane placed at $\wr = 0$. Our objective is to dimensionally reduce along the $\wr$-direction. We use $\bar{i}, \bar{j}$ to run over the spatial directions excluding $\wr$. We use $\bar{a}, \bar{b}$ to run over the spacetime directions excluding $\wr$. Here we are adopting the notation of \cite{Aharony:2003qf} and $\wr$ here should be equated with $r$ in \cite{Aharony:2003qf}. (See Figure \ref{figdimred}.)

We will focus on warped geometries where $\bar{N}$ is independent of $\wr$ and consider a background solution of the vacuum Einstein equations where $\bar{\gamma}_{\bar{i} \bar{j}} = \bar{g}_{\bar{i} \bar{j}}$. We then look for nearby solutions where,
\be
\label{warpedperturbation}
\bar{\gamma}_{\bar{i} \bar{j}} = \bar{g}_{\bar{i} \bar{j}} + \bar{h}_{\bar{i} \bar{j}}.
\ee

Note that the perturbation displayed in \eqref{warpedperturbation} is not the most general perturbation of the warped geometry. In general the higher-dimensional graviton reduces, upon dimensional reduction in the $\wr$-coordinate, to a scalar, a vector and a lower-dimensional graviton. We focus only on the lower-dimensional graviton and neglect the scalar and vector modes.

In terms of the perturbation defined in \eqref{hdefn} we have,
\be
h_{\bar{i} \bar{j}} = e^{2 A(\wr)} \bar{h}_{\bar{i} \bar{j}}.
\ee

We denote the trace of this perturbation by $\bar{h} \equiv h^{i}_{i} = \bar{g}^{\bar{i} \bar{j}} \bar{h}_{\bar{i} \bar{j}}$, where $\bar{g}^{\bar{i} \bar{j}}$ is the inverse of $\bar{g}_{\bar{i} \bar{j}}$. Note that the warp factor drops out of this trace.

Some algebra leads to the following metric-compatible connection coefficients:
\begin{equation}
 \begin{split}
 \Gamma^{\wr}_{\bar{a}\bar{b}}&=-e^{2A(\wr)} A' (\wr) \bar{g}_{\bar{a}\bar{b}},\\
 \Gamma^{\bar{a}}_{\wr\bar{b}}&=\Gamma^{\bar{a}}_{\bar{b}\wr}=A'(\wr) \delta^{\bar{a}}_{\bar{b}},\\
 \Gamma^{\bar{a}}_{\bar{b}\bar{c}}&=\bar{\Gamma}^{\bar{a}}_{\bar{b}\bar{c}}.\label{eq:connection}
 \end{split}
\end{equation}
where $\bar{\Gamma}$ refers to the connection coefficients of the metric $\bar{g}_{\bar{a} \bar{b}}$. Therefore, for any spatial vector field $J^{i}$ we have,
\begin{equation}
 \nabla_i J^i =\partial_\wr J^\wr +A'(\wr) (d-1) J^\wr +\bar{\nabla}_{\bar{i}}J^{\bar{i}},
\end{equation}
where $\bar{\nabla}_{\bar{a}}$ is the $\bar{g}_{\bar{a}\bar{b}}$-compatible covariant derivative.
Then from the definition \eqref{conservcurrentdef}, a straightforward calculation shows that,
\begin{equation}
J^\wr= -\bar{N} e^{A(\wr)} \partial_{\wr}\bar{h},
\end{equation}
where we have used the fact that $\bar{N}$ is independent of $\wr$. 
Hence we have,
\begin{equation}
\nabla_i J^i= -e^{-(d-1) A(\wr)} \bar{N} \partial_\wr \left(e^{d A(\wr)} \partial_\wr \bar{h} \right) + \bar{\nabla}_{\bar{i}} J^{\bar{i}}.
\end{equation}

The natural modes $\phi_n$ in the $\wr$-direction are those that satisfy
\be
\label{massmodes}
-e^{-(d-1) A(\wr)} \partial_\wr \left(e^{d A(\wr)} \partial_\wr \phi_n \right) = e^{-A(\wr)} m_n^2 \phi_n,
\ee
where $m_n$ gives the spectrum of the masses of the KK descendants of the graviton. This is the same eigenvalue equation that was found to govern the spectrum of massive transverse traceless graviton fluctuations in this system and it is well known that the eigenvalues $m_n^2$ are all nonzero \cite{Karch:2000ct}. Here we use these same eigenvalues in the constraint equations for the ``T'' component of the metric fluctuation.

It is natural to expand,
\be
\bar{h} = \sum \bar{h}_n \phi_n.
\ee

In the equation above, it is understood that $h_n$ varies only along the brane directions and the $\wr$-dependence is completely captured by $\phi_n$. We will use the same convention for other decompositions below.

It is natural to also define a lower-dimensional current,
\be
\bar{J}^{\bar{i}} = e^{A(\wr)} J^{\bar{i}},
\ee
and lower-dimensional energy density,
\be
\bar{\rho} = e^{2 A(\wr)} \rho.
\ee

The powers of the warp-factor that appear above can be obtained by carefully keeping track of the relative factors that appear if one uses the lower-dimensional metric $\bar{g}_{\bar{i} \bar{j}}$ to define a lower-dimensional Hamiltonian density and the lower-dimensional perturbation $\bar{h}_{\bar{i} \bar{j}}$ in \eqref{conservcurrentdef} to define a lower-dimensional current.

By also expanding the divergence of this lower-dimensional current and energy-density in terms of the modes \eqref{massmodes},
\be
\bar{J}^{\bar{i}} = \sum \bar{J}_n^{\bar{i}} \phi_n; \qquad \bar{\rho} = \sum \bar{\rho}_n \phi_n,
\ee
we find from \eqref{constraintcurved} the constraint equation,
\be
\label{warpedmassiveconstraint}
m_n^2 \bar{N} \bar{h}_n +\bar{\nabla}_{\bar{i}}\bar{J}_n^{\bar{i}} = 16 \pi G \bar{N} \bar{\rho}_n.
\ee

We again see that this equation allows for a nonzero matter stress-energy that cannot necessarily be measured at the boundary of the brane because of the $m_n^2$ term. Using this term, it is possible to find a solution to the constraint \eqref{warpedmassiveconstraint} where the value of $\bar{\rho}_n$ is not equal to the divergence of $\bar{J}_n^{\bar{i}}$ and the difference is made up by the $m_n^2$ term.

The underlying physics can be understood as follows. In the higher-dimensional setting, it is clear that the constraint equation can be satisfied by shooting the gravitational field lines ``off of'' the brane and allowing them to end at the asymptotic boundary. In the lower-dimensional theory, this effect is captured by $m_n^2$ term. This is because the $m_n^2$ term arises directly from the variation of the metric in the directions that point ``off'' the brane. Therefore, the nonzero $m_n^2$ term is a signal in the lower-dimensional effective field theory that the gravitational field lines can escape off the brane in the higher-dimensional description.

As we have already mentioned, it is known through direct calculation that $m_n^2$ is nonzero for AdS branes embedded in AdS spaces. But even if this had not been known, our analysis above could have been used to deduce this fact.

We would like to make a few comments.
\begin{enumerate}
\item
In a previous paper \cite{Geng:2020fxl}, we studied the system with two branes, where a specific linear combination of the localized gravitons is massless. One of the results of \cite{Geng:2020fxl} was that if one studies minimal surfaces with endpoints on both branes then the only such surface is the horizon. In the higher-dimensional description, this implies that the entanglement wedge of the defect, where the branes meet, is the entire exterior of the black hole.  Therefore, even in the dimensionally-reduced description, this entanglement wedge extends up to the asymptotic boundary and is not an island. In \cite{Geng:2020fxl}, we also studied islands on a single brane that are relevant for the entanglement between internal degrees of freedom on the defect. 
But if one studies islands on one brane then the effective description in terms of the localized-gravity theory on that brane has access only to a linear combination of a massive and massless graviton. Correspondingly, the asymptotic form of the metric on a single brane does not give us access to the total energy on that brane. Therefore the results of \cite{Geng:2020fxl} are entirely consistent with the reasoning advanced in this paper.
\item
It is sometimes proposed that an island with a massless graviton can be studied by starting with an AdS black hole that partially evaporates into a non-gravitational bath allowing for a massive graviton, after which one ``switches off'' the coupling between the bath and the AdS space. However, a closer analysis of the causal structure of this process reveals that the graviton is always massive in the island. This is explained in appendix \ref{appdecoupling}.
\item
It is also not possible to obtain operators that commute with the asymptotic Hamiltonian by dressing them to an ``end-of-the-world'' brane, since such a brane must interact with the ambient metric. Therefore, in theories with long-range gravity, the brane must itself be dressed to the asymptotic boundary. The asymptotic boundary Hamiltonian then measures the energy of the brane and that of additional excitations. So even if one dresses an excitation to the brane, the excitation still transforms under time-translations generated by the asymptotic Hamiltonian.
\end{enumerate}

\section{Islands in decoupled systems \label{secjt}}

In the sections above, we described a puzzle and then showed how the puzzle could be resolved if the gravitational theory that supports islands is massive. When a standard theory of gravity in AdS is coupled to a bath, the mass of the graviton arises as a consequence of the coupling.

Islands have also been studied when a gravitating system and another system are entangled but not coupled \cite{Penington:2019kki,Hartman:2020khs,Balasubramanian:2020coy,Anderson:2021vof,Fallows:2021sge,Miyata:2021ncm}. In higher dimensions, islands in decoupled systems are also subject to the puzzle described in section \ref{secpuzzle}. In the absence of a coupling, the graviton does not pick up a dynamically generated mass, and so the resolution outlined in section \ref{secmassiveresolve} does not apply to decoupled systems. Therefore, in higher dimensions, unless the theory of gravity is massive to start with, our analysis above implies that islands in decoupled systems would suffer from an inconsistency with the Gauss law.

However, most of the studies of decoupled islands have been performed only in $1+1$-dimensions and so the analysis of section \ref{secpuzzle} and section \ref{secmassiveresolve} does not directly apply to these studies. Although it will be of interest to see if the considerations above apply to this lower-dimensional case, 
in this section, we focus on another elementary consistency condition that must be obeyed by all islands in decoupled systems, including islands in $1+1$-dimensional theories. We will explain how this simple separate consistency condition--though not necessarily ruling out islands--is sufficient to indicate that islands in decoupled systems cannot be used to model realistic evaporating black holes.

\subsection{Consistency condition}
Consider two decoupled systems described by Hilbert spaces, ${\cal H}_1$ and ${\cal H}_2$ respectively. The joint system is described by the Hilbert space,
\be
\label{jointhilb}
{\cal H}={\cal H}_1 \otimes {\cal H}_2.
\ee

That the systems are {\em decoupled} means, by definition, that the Hamiltonian in the joint Hilbert space $H$ is just a sum of the Hamiltonians $H_1$ and $H_2$,
\be
\label{jointhamilt}
H = H_1 \otimes 1 + 1 \otimes H_2, \qquad \text{(decoupled~systems)}
\ee

We now derive a few consequences of the elementary equations \eqref{jointhamilt} and \eqref{jointhilb}. First note that given any density matrix $\rho$ that describes the state of the joint system, we can obtain a density matrix for system $1$ using
\be
\rho_1 = \tr_2(\rho).
\ee

Now consider the action of an {\em arbitrary unitary operator} $U_2$ that acts on ${\cal H}_2$, i.e. it acts as the identity on ${\cal H}_1$. This modifies the joint density matrix as
\be
\rho \rightarrow U_2 \rho U_2^{\dagger}.
\ee

But under such a transformation of state, the density matrix of the first system is unchanged,
\be
\label{commutop}
\tr_2 (U_2 \rho U_2^{\dagger}) = \tr_2(\rho) = \rho_1.
\ee

Consequently, the expectation value of {\em all} observables in system 1 are unchanged by the action of a unitary in system 2. 

Equation \eqref{commutop} holds whenever the Hilbert space factorizes. But in the case of decoupled systems, as a consequence of \eqref{jointhamilt}, even the time-evolution operator in system $2$ commutes with all operators in system $1$. Therefore \eqref{commutop} must hold even if we evolve $U_2$ forward or backward by an {\em arbitrary} amount of time $t$. 
\be
\label{commutimeevol}
\tr_2 \left[e^{-i (1 \otimes H_2) t} U_2 e^{i (1 \otimes H_2) t} \rho e^{-i (1 \otimes H_2) t} U_2^{\dagger} e^{i (1 \otimes H_2) t}\right] = \tr_2(\rho) = \rho_1.
\ee

Now consider a setup where we have two decoupled systems, and where a region $R$ in the second system is believed to also describe the degrees of freedom in an island $I$. The argument above tells us that no possible unitary operator acting on $R$, at any point in time, should have the ability to affect any degrees of freedom from the first system. Conversely no possible unitary acting on the first system at any point of time should affect the island. This leads us to the following elementary consistency condition.
\begin{observation}
When islands are redundant with degrees of freedom from one part of a decoupled pair of systems, it should be impossible for the islands to either send signals to or receive signals from the degrees of freedom described by the other part of the decoupled pair of system.
\end{observation}

\subsection{Implications of the consistency condition}
We now explain how the consistency condition derived above rules out the possibility of modeling the interior of physical black holes using islands in decoupled systems.

\begin{figure}[!ht]
\begin{center}
\includegraphics[width=0.7\textwidth]{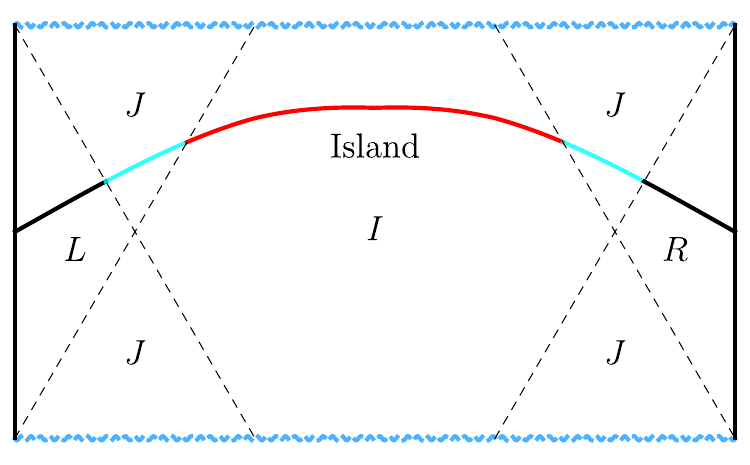}
\caption{\em One possibility for an island in a decoupled pair of systems. The island cannot affect the degrees of freedom beyond either the left horizon or the right horizon. The solid black lines are the left and right asymptotic AdS boundaries. The broken blue lines are singularities and the dashed lines are horizons. The dashed lines also demarcate the boundaries between regions marked, $L$, $R$, $I$, $J$. In this figure, the horizontal direction is spatial and time runs along the vertical direction. \label{figdecoupled}}
\end{center}
\end{figure}

An example of a geometry that obeys the consistency condition above is provided by Figure \ref{figdecoupled}, which was obtained in \cite{Balasubramanian:2020coy}.\footnote{The setup of \cite{Hartman:2020khs} also satisfies our consistency condition although this is more subtle to see from the Penrose diagrams of \cite{Hartman:2020khs}. In Figure 13 of \cite{Hartman:2020khs} for instance, the independent degrees of freedom of the gravitational system should be associated only with the ``tip'' of the Penrose diagram at spatial infinity, and one can neither send nor receive signals from spatial infinity. We thank Tom Hartman for explaining this point to us.} Note that in Figure \ref{figdecoupled} it is insufficient for the island to be behind the future horizon to satisfy the consistency condition. Starting from the left asymptotic boundary, one has to cross {\em two} horizons to reach the island region, and the same is true if one starts from the right asymptotic boundary.

Referring to Figure \ref{figdecoupled}, the two asymptotic boundaries completely describe the physics in region $L$ and region $R$. The island is dual to a decoupled system that completely describes the physics in region $I$. The regions marked $J$ are described by a combination of the decoupled system and the system that lives on the asymptotic boundaries. This Penrose diagram meets our consistency condition since every point in region $L$ and region $R$ is separated by a spacelike interval from every point in region $I$. Note that signals sent from region $L$ (or region $R$) and region $I$ can meet in the regions marked $J$. This does not contradict the consistency condition but reflects the familiar feature, also present in the duality between an eternal black hole and the thermofield doubled state \cite{Israel:1976ur,Maldacena:2001kr}, that signals from decoupled systems can meet inside a wormhole.

\begin{figure}[!ht]
\begin{center}
\includegraphics[width=0.5\textwidth]{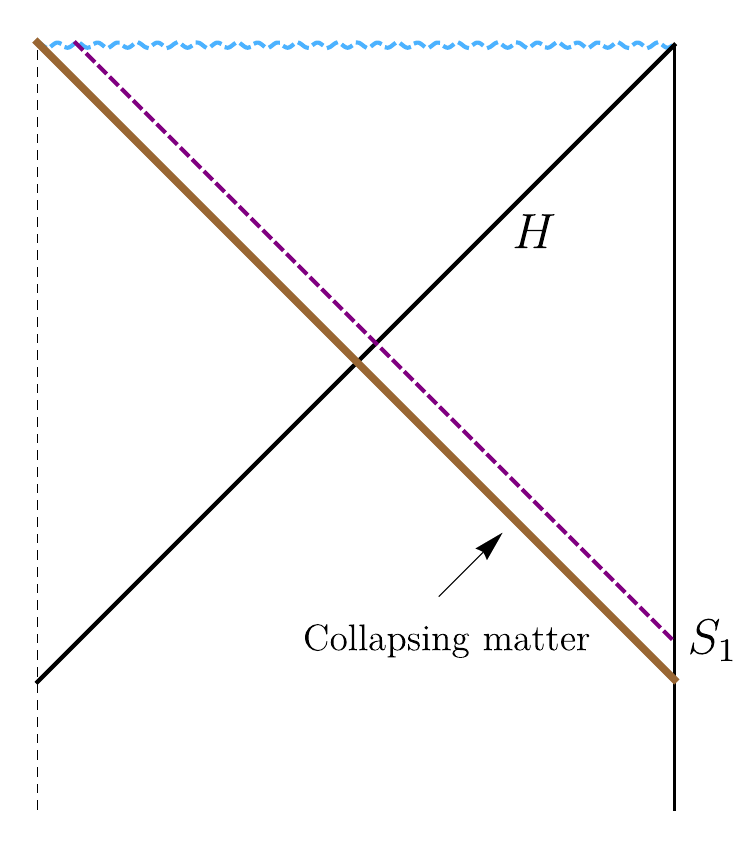}
\caption{\em A single sided black hole formed from the collapse of some matter (shown in brown). The dashed line on the left is the origin of polar coordinates and the solid black line on the right is the asymptotic AdS boundary. The Figure shows a signal $S_1$ that originates on the boundary and reaches deep inside the interior.  In this figure, the horizontal direction is spatial and time runs along the vertical direction. \label{figsingle}}
\end{center}
\end{figure}

This example illustrates that any spacetime geometry that meets the above consistency condition must have a qualitatively different causal structure from that of a black hole formed from collapse. For contrast, a Penrose diagram of a single-sided AdS black hole is shown in Figure \ref{figsingle}. It is clear that, even classically, it is possible to send signals to {\em every} point in the interior of the black hole from the exterior, provided the signal is sent early enough. If any part of the interior had been part of an island from a decoupled system, this would not have been possible by our consistency condition. Therefore no part of the interior of such a black hole can constitute an island that is described by degrees of freedom from a decoupled system.

Figure \ref{figsingle} shows a large single-sided black hole, but the same conclusion holds for an evaporating black hole. 

\begin{figure}[!ht]
\begin{center}
\includegraphics[width=0.5\textwidth]{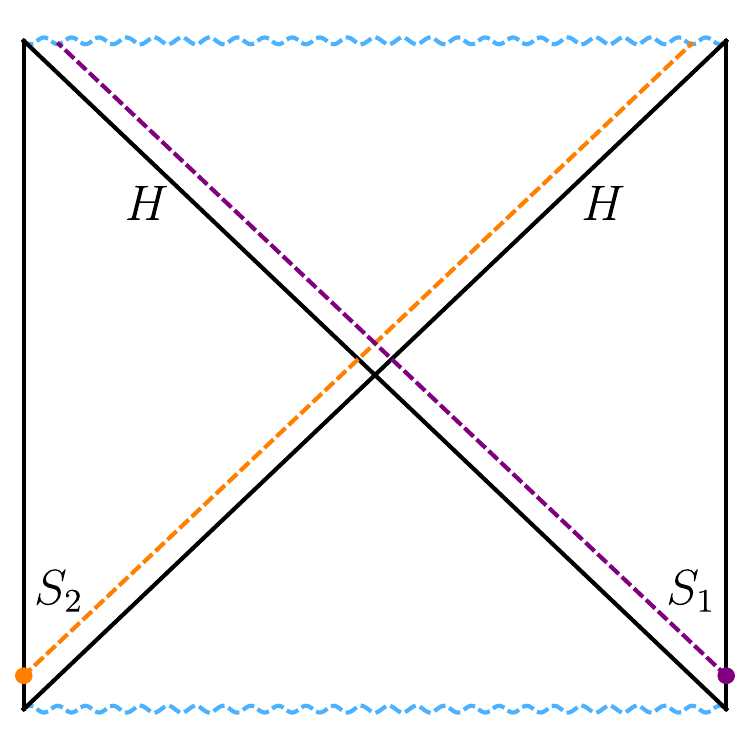}
\caption{\em In the eternal black hole, it is always possible to send signals to any point in the interior, even classically, provided these signals originate at an early enough time. The Figure shows a signal $S_1$ that originates on the left boundary and another signal $S_2$ that originates on the right boundary. Therefore, no part of the interior of a standard eternal black hole can constitute an island that is redundant with degrees of freedom from a decoupled system. In this figure, the horizontal direction is spatial and time runs along the vertical direction. \label{figeternal}}
\end{center}
\end{figure}

For completeness, we mention that the Penrose diagram in Figure \ref{figdecoupled} is also qualitatively different from that of the standard eternal black hole, which is shown in Figure \ref{figeternal}. Here, it is always possible to send a signal to any point in the interior from either the left or the right asymptotic boundary. The consistency condition above then implies that no part of the interior of the eternal black hole of Figure \ref{figeternal} can constitute an island corresponding to degrees of freedom from a system that is decoupled from both asymptotic boundaries.

We emphasize that although the Penrose diagram in Figure \ref{figdecoupled} satisfies the consistency condition described in this section, this picture of an island is not exempt from the analysis described in section \ref{secpuzzle}. In higher dimensions, Figure \ref{figdecoupled} would be consistent for massive gravity but would violate the Gauss law in ordinary massless gravity since the energy of excitations in the island could be measured by combining observations in the left and right asymptotic regions. The study in \cite{Balasubramanian:2020coy}, where this diagram was obtained, was performed in the $1+1$ dimensional context of JT gravity, where we have yet to determine if some analogue of our puzzle exists.

Islands in the presence of a decoupled bath were also studied in \cite{Penington:2019kki}.
The theory used in \cite{Penington:2019kki} is defined via a precise rule for Euclidean path integrals. We are unable to determine if the Euclidean saddle points corresponding to islands found in \cite{Penington:2019kki} satisfy the consistency condition outlined here, since the Lorentzian action that gives rise to the rules for the Euclidean path integral above is not known. It would be interesting to obtain the Lorentzian theory corresponding to the Euclidean rules of \cite{Penington:2019kki}. This would allow us to check if the consistency condition outlined in this section holds for this theory and would more readily allow an extension of the puzzle of section \ref{secpuzzle} to this model.

\section{Discussion \label{secdiscussion}}
In this paper, we have described how models of black hole evaporation that involve ``islands'' lead to puzzles in theories with long-range gravity where the Gauss law applies. In quantum mechanics, the Gauss law is stronger than it is in classical mechanics. A localized operator in the bulk must have nonzero energy, and so it must fail to commute with the Hamiltonian which, in a theory of gravity, is a boundary term. This nonzero commutator can be thought of as arising because any localized operator must be ``dressed'' to the asymptotic boundary. Such a dressing is sometimes called a gravitational Wilson line. (See \cite{Donnelly:2016rvo,Giddings:2019hjc} and references there.)

Since there are no negative charges in gravity, gravitational Wilson lines end only on asymptotic boundaries. In conventional island-free entanglement wedges in AdS/CFT, such asymptotic regions are always part of the wedge. Therefore operators in the wedge can form a self-contained algebra, since the entire operator, including its dressing, can be localized within the wedge. 

Islands would be entanglement wedges that do not extend to the asymptotic boundary and are surrounded by their complements. Consequently, in an ordinary theory of gravity in which the Gauss law applies, it is not possible to localize an operator and its dressing entirely to an island. This means that, in an ordinary gravitational theory, an island cannot constitute an entanglement wedge.

We showed how this puzzle was resolved in a concrete setting. When islands are realized by embedding a brane in a higher-dimensional anti-de Sitter spacetime, the lower-dimensional theory of gravity obtained on the brane has a massive graviton. Such theories do not have a Gauss law, and so the puzzle does not arise at all.

An extension of our analysis to $1+1$ dimensional theories of gravity would allow contact with the significant literature where islands have been studied using JT gravity \cite{Almheiri:2019qdq,Penington:2019kki} and other $1+1$ dimensional models \cite{Hartman:2020swn,Akal:2020twv,Gautason:2020tmk}. One obstacle to extending our analysis is purely technical: the analysis of gravitational constraints reviewed in \cite{Raju:2020smc} which suggests that gravity stores information differently from local quantum field theories is valid only for spacetime dimension larger than two. Moreover, the mechanism whereby imposing transparent boundary conditions on a theory of gravity coupled to matter in AdS leads to a mass for the graviton is understood only for higher-dimensional theories \cite{Porrati:2003sa,Porrati:2001gx,Porrati:2002dt}.

Nevertheless, we would like to mention some results in the existing literature that suggest that it might be possible to generalize our puzzle.
Even in $1+1$ dimensional dilaton-gravity models with a negative cosmological constant and appropriate boundary conditions, it is possible to define conserved charges \cite{Grumiller:2017qao,Grumiller:2007ju,Ruzziconi:2020wrb} that are related to the constraints of the theory. Moreover \cite{Harlow:2018tqv}, extending the techniques of \cite{Harlow:2015lma,Guica:2015zpf}, analyzed the significance of the bulk constraints in pure JT gravity without matter and showed that these constraints prevent the factorization of the Hilbert space on the boundary, which is a puzzle that is somewhat similar in flavor to the puzzle that we have described in this paper. Some further discussion of these issues can be found in \cite{Gao:2021uro,Harlow:2020bee}. 
One aspect of the higher-dimensional analysis that clearly does carry over even to $1+1$ dimensions is the fact that coupling to an external bath will allow energy to leak out of the system and thereby ruin any conservation laws that were present before the coupling to the bath. It is an interesting open problem to build on these papers and see whether and how our puzzle extends to $1+1$ dimensions. 

Similarly, we also point out that we have addressed only the theories of massive gravity that emerge when AdS branes are embedded in AdS spaces. Our analysis does not necessarily indicate whether other theories of massive gravity can support islands. 

Finally we contrast the picture of how quantum information is localized in theories of gravity that support islands with the corresponding picture in theories of long-range gravity. Even in theories with long-range gravity, degrees of freedom that appear to be localized to a region may be equated with a scrambled version of degrees of freedom in another region. This physical effect follows from a careful analysis of gravitational constraints as reviewed in \cite{Raju:2020smc}. It is also important for black hole evaporation and for understanding the interior of black holes in AdS/CFT as emphasized in \cite{Papadodimas:2012aq,Papadodimas:2013jku}. 
However, in theories with long-range gravity, degrees of freedom near the asymptotic boundary capture {\em all} the degrees of freedom on a bulk Cauchy slice and not only the degrees of freedom in an island. Therefore, in the presence of long-range gravity, information about the black hole is always available outside, regardless of the stage of black-hole evaporation \cite{Laddha:2020kvp}.

On the other hand, in models of black hole evaporation that involve islands, information emerges gradually during black hole evaporation according to the Page curve. This agrees with intuition obtained from non-gravitational systems, where such a Page curve is expected on very general grounds. They key reason that the entropy of the radiation first increases and then decreases in such models is that the radiation region describes the island but does not describe its complement. But not only do such models necessarily involve non-gravitational baths; they seem to be consistent only in theories without a Gauss law. 

We emphasize that our puzzle is not simply a question of the examples presented in the literature. Theories with long-range gravity allow the energy to be measured at the asymptotic boundary, which is in contradiction with the notion of an island as a connected component of an entanglement wedge, disconnected from the boundary.
This is why all consistent constructions of islands so far in gravitational theories with more than two dimensions involve massive gravity.

\section*{Acknowledgments}
We are grateful to Chris Akers, Ahmed Almheiri, Tuneer Chakraborty, Joydeep Chakravarty, Chandramouli Chowdhury, Victor Godet, Daniel Grumiller, Tom Hartman, Chethan Krishnan, Alok Laddha, Hong Liu, Juan Maldacena, Henry Maxfield, Ruchira Mishra, Kyriakos Papadodimas, Olga Papadoulaki, Priyadarshi Paul, Massimo Porrati, Siddharth Prabhu and Pushkal Shrivastava for helpful discussions. We are grateful to Victor Godet, Daniel Harlow, Daniel Jafferis, Raghu Mahajan and Massimo Porrati for helpful comments on a draft of this manuscript. HG is very grateful to his parents and recommenders. The work of AK and MR was supported, in part, by a grant from the Simons Foundation (Grant 651440, AK). The work of CP was supported in part by the National Science Foundation under Grant No.~PHY-1914679 and by the Robert N. Little Fellowship. The work of SR was partially supported by a Swarnajayanti fellowship, DST/SJF/PSA-02/2016-17 from the Department of Science and Technology (India). Research at ICTS-TIFR is supported by the government of India through the Department of Atomic Energy grant RTI4001. The work of LR is supported by NSF grants PHY-1620806 and PHY-1915071, the Chau Foundation HS Chau postdoc award, the Kavli Foundation grant ``Kavli Dream Team,'' and the Moore Foundation Award 8342. SS was supported by National Science Foundation (NSF) Grants No. PHY-1820712 and PHY-1914679.

\appendix

\section*{Appendix}

\section{Loopholes in possible counterarguments  \label{appalternatives}}

In the main text we have shown that islands cannot support approximately local operators that commute with operators in its complement. 

In this appendix, we will examine some proposals that attempt to sidestep our puzzle. 
We will conclude, as the puzzle in this paper would lead us to expect, that none of these proposals lead to operators that are viable for entanglement wedge reconstruction in a standard theory of gravity. 

\subsection{Products of modes in frequency space}

It is not difficult to construct operators that commute with the Hamiltonian to an excellent approximation. As explained in the text, the difficulty is in localizing such operators. This follows from the uncertainty principle; an operator that is localized to a finite region must have nonzero energy and an operator that has zero energy must extend to infinity.  

This can be illustrated using the following example.\footnote{We thank an anonymous referee for drawing our attention to this proposal.} As in section \ref{secpuzzle}, consider a scalar field $\phi$ propagating in a spherically symmetric black hole in an asymptotically global \ads[\dt+1] spacetime. In this subsection, we write the field as  $\phi(t, r, \Omega)$, thereby explicitly specifying its position in time, the radial coordinate and the $S^{\dt-1}$. 
To leading order in $G$, outside the black hole horizon, the field can be expanded in terms of modes,
\be
\label{phiexpansion}
\phi(t, r, \Omega) = \sum_{\ell} \int d \omega\, a_{\omega, \ell} e^{-i \omega t} \psi_{\omega, \ell}(r) Y_{\ell}(\Omega) + \text{h.c},
\ee
where $Y_{\ell}(\Omega)$ is a spherical harmonic corresponding to the angular momentum quantum numbers $\ell$ and $\psi_{\omega, \ell}$ are radial mode functions that are discussed in greater detail in \cite{Papadodimas:2012aq}.  With respect to the boundary Hamiltonian $H$ that appeared in section \ref{secpuzzle} as the integral of the boundary metric we have,
\be
\label{boundh}
[H, a_{\omega, \ell}] = -\omega a_{\omega, \ell}; \qquad [H, a_{\omega, \ell}^{\dagger}] = \omega a_{\omega, \ell}^{\dagger}; \qquad [a_{\omega, \ell}, a_{\omega', \ell'}^{\dagger}] = \delta(\omega - \omega') \delta_{\ell \ell'}.
\ee
Now consider the following operator,
\be
X = a^{\dagger}_{\omega_1,0} a_{\omega_2,0} a_{\omega_3,0},
\ee
where $\omega_1 = \omega_2 + \omega_3$ and where the $0$ in the subscript indicates that we are focusing on the $s$-wave sector.  It is easy to see from \eqref{boundh} that 
\be
[H, X] = 0.
\ee
The trilinear operator $X$ above is just an example and the discussion below easily generalizes to any polynomial in the modes that commutes with the Hamiltonian.

However, the creation and annihilation operators that enter inside $X$ are not local operators. As a result, whereas $X$ commutes with the Hamiltonian it does not commute with other operators outside the island. We will now show, through an explicit computation, that the nonzero commutator between $X$ and operators outside the island can be seen at $\Or[1]$ and does not require us to even keep track of gravitational effects.

Consider the limit of the field operator near the boundary at a time $t_1$. We can take $t_1$ to be sufficiently small so that the boundary point is spacelike to every point in the island. Note that such points always exist since, by our definition in section \ref{definition}, the island does not extend to the boundary.  Using the mode expansion and commutation relations above we find that (to $\Or[G^0]$),
\be
\label{nonzerocomm}
\lim_{r \rightarrow \infty} r^{\Delta} [X, \phi(r, t_1, \Omega)] = -C_{\omega_1} e^{-i \omega_1 t_1} a_{\omega_2, 0} a_{\omega_3,0} + C_{\omega_2} e^{i \omega_2 t_1} a_{\omega_1, 0}^{\dagger} a_{\omega_3,0} +  C_{\omega_3} e^{i \omega_3 t_1} a_{\omega_1, 0}^{\dagger} a_{\omega_2,0},
\ee
where we have normalized the spherical harmonics through $Y_{0}(\Omega)=1$ and $C_{\omega}$ is a real number defined by 
\be
C_{\omega} = \lim_{r \rightarrow \infty} r^{\Delta} \psi_{\omega,0}(r).
\ee
The precise value of $C_{\omega}$ must be computed numerically in general $\dt$ but will not be required below.

The nonzero commutator \eqref{nonzerocomm} can be easily detected inside a correlation function by inserting two more field operators, at points $(r, t_2, \Omega_2)$ and $(r, t_3, \Omega_3)$ that we again choose to be spacelike to the island, and using the correlators 
\be
\label{thermalcorr}
\langle  \Psi | a_{\omega, \ell} a_{\omega', \ell'}^{\dagger} |\Psi \rangle = G^+_{\omega}  \delta(\omega - \omega') \delta_{\ell, \ell'}; \qquad 
\langle \Psi| a_{\omega, \ell}^{\dagger} a_{\omega', \ell'} |\Psi \rangle = G^{-}_{\omega}  \delta(\omega - \omega') \delta_{\ell, \ell'},
\ee
where 
\be
G^{+}_{\omega} = {1 \over 1 - e^{-\beta \omega}}; \qquad G^{-}_{\omega} = G^{+}_{\omega} - 1 = {e^{-\beta \omega} \over 1 - e^{-\beta \omega}},
\ee
and $\beta$ is the inverse temperature of the black hole. We find that
\be
\label{xcomm}
\begin{split}
&\lim_{r \rightarrow \infty} r^{3 \Delta} \langle \Psi| \phi(r, t_2, \Omega_1) \phi(r, t_3, \Omega_2)   [X, \phi(r, t_1, \Omega)] |\Psi \rangle =   C_{\omega_1} C_{\omega_2} C_{\omega_3} \\ &\times \left[  -e^{-i \omega_1 t_1}G^{-}_{\omega_2} G_{\omega_3}^{-}\left(e^{i \omega_2 t_2 + i \omega_3 t_3} + e^{i \omega_3 t_2 + \omega_2 t_3} \right)  +   e^{i \omega_2 t_1} G^{+}_{\omega_1} G_{\omega_3}^{-}  \left(e^{-i \omega_1 t_2 + \omega_3 t_3} + e^{-i \omega_1 t_3 + \omega_3 t_2} \right)  \right. \\ &\left. +    e^{i \omega_3 t_1} G^{+}_{\omega_1} G^{-}_{\omega_2} \left(e^{-i \omega_1 t_2 + \omega_2 t_3} + e^{-i \omega_1 t_3 + \omega_2 t_2} \right) \right].
\end{split}
\ee
Note that once we use $\omega_1 = \omega_2 + \omega_3$ we see that the result depends only on the time-differences $(t_2 - t_1)$, $(t_3 - t_1)$ and $(t_3 - t_2)$.

We can present this result in a manner that is identical to section \ref{secpuzzle} by showing how a unitary operator containing $X$ changes correlators outside the island. $X$ is not Hermitian, but we can construct a unitary operator via $U_X = e^{i \lambda (X + X^{\dagger})}$. We then find that
\be
\label{unitaryexpec}
\begin{split}
&\lim_{r \rightarrow \infty} r^{3 \Delta} \left. {\partial \over \partial \lambda} \langle \Psi| \phi(r, t_1, \Omega_1) \phi(r, t_2, \Omega_2)   U_X \phi(r, t, \Omega) U_X^{\dagger} |\Psi \rangle \right|_{\lambda = 0}
\\ &= \lim_{r \rightarrow \infty} i r^{3 \Delta} \langle \Psi| \phi(r, t_1, \Omega_1) \phi(r, t_2, \Omega_2)   [X+X^{\dagger}, \phi(r, t_1, \Omega)] |\Psi \rangle  \neq 0.
\end{split}
\ee
The exact expression on the right hand side above can be read off from  \eqref{xcomm} but we do not write it since it is not illuminating.

We pause to note another subtlety with the unitary $U_X$. One might have thought that the unitary changes the occupation number of modes with frequency $\omega_1, \omega_2$ and $\omega_3$. However, a simple check shows that the correlators \eqref{thermalcorr} are unchanged in the state $U_X|\Psi \rangle$ for all frequencies including these three frequencies. More precisely,
\be
\langle \Psi| U_X^{\dagger} a_{\omega} a_{\omega}^{\dagger} U_X |\Psi \rangle = \langle \Psi| a_{\omega} a_{\omega}^{\dagger} |\Psi \rangle, \quad \text{and} \quad  \langle \Psi| U_X^{\dagger} a^{\dagger}_{\omega} a_{\omega} U_X |\Psi \rangle = \langle \Psi| a_{\omega}^{\dagger} a_{\omega}|\Psi \rangle, \qquad \forall \omega
\ee
This is a general property of unitary operators that commute with the Hamiltonian. They do not ``create excitations'' in typical states because correlators in typical states are thermal correlators, and so if $A$ is any operator,
\be
\langle \Psi | U_X^{\dagger} A U_X |\Psi \rangle = {1 \over Z(\beta)} \tr(e^{-\beta H} U_X^{\dagger} A U_X) = {1 \over Z(\beta)} \tr(U_X e^{-\beta H} U_X^{\dagger} A) = \langle \Psi | A |\Psi \rangle,
\ee
where we have used the cyclicity of the trace and the commutator of $U_X$ and the Hamiltonian, and $Z(\beta)$ is the partition function.

So $U_X |\Psi \rangle$ is not an excited state at all and can be thought of as just another black hole microstate. This observation is not in contradiction with \eqref{unitaryexpec} since the unitary operator there is inserted in the middle of other operators.

To summarize, our calculation implies that $X$ cannot be thought of, in any sense, as an operator localized to the island. Moreover, the action of a unitary made up of $X$ does not even create an excitation. So its existence does not resolve the puzzle presented in the text and, moreover, is irrelevant for the discussion of entanglement wedge reconstruction.

The discussion above is quite general and can be generalized to all operators that can be formed from the operators displayed in \eqref{phiexpansion}. We note that by using state-dependent operators it is possible to construct operators that cannot be detected by means of \eqref{xcomm} \cite{Papadodimas:2015jra} but even these operators are infinitely delocalized  and are not relevant for entanglement wedge reconstruction. 

\subsection{Swapping excitations}
Another proposal is to only study operators that swap one excitation for another. More specifically, one starts not with an empty black hole but rather with an excited black hole state of the form $U |\Psi \rangle$ where $U$ is defined in section \ref{secpuzzle}. Now imagine that one has another field with precisely the same mass, which we denote by $\tilde{\phi}$. One might then consider the operator that {\em swaps} an excitation made up of $\phi$ for an excitation made up of $\tilde{\phi}$. More precisely we study the unitary operator $V = \tilde{U} U^{\dagger}$ where $\tilde{U} = e^{i \lambda \tilde{\phi}(P)}$ in the notation of section \ref{secpuzzle}.  

We note that to leading order in $G$,
\be
\lim_{r \rightarrow \infty} r^{\dt-2 + \Delta} \int_{(\partial\text{AdS})} {\partial \over \partial \lambda} \left. \langle \Psi | \tilde{U} {\partial \phi (P')  \over \partial t }  h_{00} \tilde{U}^{\dagger} | \Psi \rangle \right|_{\lambda = 0} = 0.
\ee
But the insertion sandwiched between the unitaries above is precisely the same insertion that appeared in section \ref{secpuzzle} where we found a nonzero value. Therefore the insertion of the boundary Hamiltonian and a matter field can distinguish between the state with an excitation of $\phi$ and the state with an excitation of $\tilde{\phi}$. Therefore the operator $V$ does not commute with the product of the boundary Hamiltonian and a matter field at infinity. 

This should not be puzzling since the correlator we are measuring keeps keep track of more than just the energy; the insertion of $\phi$ breaks the symmetry between the two states. Indeed, one can check that for a different correlator,
\be
\lim_{r \rightarrow \infty} r^{\dt-2 + \Delta} \int_{(\partial\text{AdS})} {\partial \over \partial \lambda} \left. \langle \Psi | \tilde{U} {\partial \tilde{\phi} (P')  \over \partial t }  h_{00} \tilde{U}^{\dagger} | \Psi \rangle \right|_{\lambda = 0} \neq 0,
\ee
and the nonzero value is precisely the value that was explored in section \ref{secpuzzle}.

\subsection{Background fields as coordinates}
In the presence of suitable background fields, it is possible to define perturbatively gauge-invariant local operators. More precisely, imagine that in $\dt+1$ spacetime dimensions we have $\dt+1$ background scalar fields $X^1 \ldots X^{\dt+1}$ and a classical background where a point in spacetime can be {\em uniquely specified} by specifying the values of the fields.  Let $\phi$ be another propagating field. It is then possible to define the operator
\be
\label{backgroundop}
O(Z^i) = \int \phi(x) F(Z^1 - X^1(x), \ldots Z^{\dt+1} -X^{\dt+1}) d x.
\ee
where $F$ is a sharply peaked function that has support only when all its arguments are close to zero.  Such an operator commutes with boundary operators within perturbation theory (although not nonperturbatively) and approximates a local operator. For more details we refer the reader to \cite{Marolf:2015jha} and references there. 

However, such a construction is not possible for a black hole at late times. By the classical no-hair theorem, at late times and in the absence of asymptotic sources, the black hole solution does not support classical fields that take on distinct values at each point in spacetime. On the other hand, if we consider asymptotic sources that stay on permanently and stabilize a background of  classical scalar hair, we would break diffeomorphism invariance and therefore give the graviton a mass as discussed in \cite{Blake:2013owa}.

Indeed, it is the emergent time-translational symmetry of the black hole solution that allows us to phrase our puzzle within perturbation theory. For general time-dependent backgrounds, it might be nonperturbatively difficult to detect the presence of a local excitation from infinity due to the existence of operators like \eqref{backgroundop}. But because the late-time geometry of the black hole is so simple, local excitations can also be detected within perturbation theory as in empty AdS.

As opposed to the eternal black hole, the time-translational symmetry is not exact for the evaporating black hole in that the evaporation of the black hole itself defines a ``clock.'' One could attempt to use this clock to define operators along the lines of \eqref{backgroundop} that commute with the boundary Hamiltonian within perturbation theory. However since the evaporation happens over a very long time scale, this clock allows one to only define  a class of very diffuse operators. These operators do not act at a well-defined time and are smeared out over an $\Or[1]$ fraction of the evaporation time scale. Such operators are very different from the approximately local operators that one seeks in standard entanglement-wedge reconstruction. However, it would be interesting to explore the graviton mass and such operators more precisely and their significance for the island proposal for evaporating black holes.

\subsection{Mundane locality}
In this paper, we have pointed out that gravitational effects pose an obstruction to defining operators that are confined to the island. The reader might wonder how this analysis is consistent with everyday experience: although we live in a world where gravity is presumably quantized, it is perfectly sensible to discuss local experiments without having to worry about the effect of these experiments at a distant location.

This has to do with the weakness of gravity as we now explain. As we have mentioned above, even in the presence of long-range gravity it is easy to obtain approximately local gauge-invariant operators. One option is simply to fix gauge.  Such gauge-fixed operators depend on the choice of gauge and obey a nonlocal algebra \cite{Donnelly:2015hta}.  But these nonlocal commutators are suppressed by a factor of $\left({E \over M_{\text{pl}}}\right)^{\dt - 2}$ where $E$ is a characteristic energy scale.  (This is also the factor that appears on the right hand side of \eqref{heisenbergeom}, where $E$ depends on the precise operators inserted in that correlator.) For the energy scales probed in ordinary experiments, this factor is well below the threshold of experimental accuracy,  and so the unusual localization of information in gravity is of no practical consequence.

On the other hand, gravitational effects are a key aspect of entanglement-wedge reconstruction. It is due to such effects that the entanglement wedge can be larger than the causal wedge, as happens when islands appear.  Therefore such effects cannot be ignored in the study of islands.

\subsection{Decoupling the bath \label{appdecoupling}}
Yet another proposal for reconstructing operators in the island in a standard theory of gravity is to start with a  theory of massive gravity coupled to a nongravitational bath, wait until an island forms in the bulk, and then turn off the coupling. One might naively expect that the graviton becomes massless in the bulk while  operators in the island are still redundant with operators in the radiation region.  Therefore a naive analysis might suggest that this procedure can be used to obtain islands in a standard theory of gravity.

We now show that this does not happen. Even in this scenario, the island forms and exists only in the presence of a massive graviton. 

In simple models, where the coupling is given by the product of a light operator on the boundary of AdS with an operator in the nongravitational bath, the coupling can be turned on and off by means of a simple change of boundary conditions at the boundary of AdS. One expects the effect of this change in boundary conditions to propagate causally in the bulk. 

\begin{figure}
\begin{center}
\includegraphics[height=0.4\textheight]{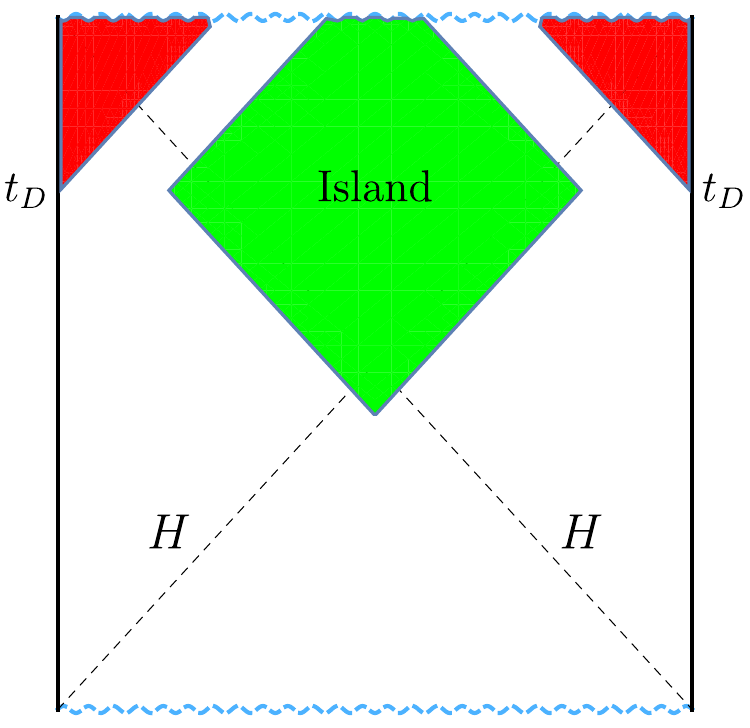}
\caption{\em A possible attempt to produce islands in standard gravity. An island is first formed in a theory of massive gravity that involves a nongravitational bath. At time $t_D$ the coupling to the bath is switched off. The change in boundary conditions is expected to make the graviton massless at all points in the causal future of the decoupling event (shown in red). But the island (shown in green) is unaffected by this event since every point in the island is spacelike to the decoupling event. So the graviton remains massive in the island. \label{figdecoupling}}
\end{center}
\end{figure}

Physically, one can think of a ``shock wave'' that propagates inwards and changes the graviton from being massive to being massless in its wake.  The graviton therefore becomes massless only at  points in the bulk that are in the {\em causal future} of the decoupling event on the boundary. This is shown in Figure \ref{figdecoupling}: the coupling is turned off in a coordinated manner at the time $t_D$ on the boundary and the graviton becomes massless in the red ``triangular'' regions on the top left and top right corner of the Penrose diagram.

We see that the entanglement wedge corresponding to the island is a causal diamond in the bulk with the property that every point in the diamond is spacelike to the decoupling event. The island is depicted as a green diamond in Figure \ref{figdecoupling}. Therefore the island never learns of the change in boundary conditions and the graviton remains massive everywhere in the island.

In the black-hole geometry, if the bath is decoupled at late times the graviton  remains massive on at least some part of every spacelike slice in the bulk.  This is a consequence of the fact that the black-hole spacetime contains nontrivial causal patches. In contrast if a bath is coupled and then decoupled in empty AdS then, after a ``transitional period'' in the bulk, the graviton becomes massless everywhere. However, even in empty AdS the island is always confined to the region where the graviton is massive.

We note that were the graviton in the vicinity of the island to indeed become massless,\footnote{We note that this possibility is the more complicated option in that it would require that the action of turning on and off the coupling is achieved by means of a  complicated operator of AdS that can, in principle, distort the geometry deep in the bulk in a noncausal fashion.} the longitudinal mode essential to the consistency with Gauss' law would disappear when the mass turns off, leading to a violation of Gauss' law even if the mass had been present at an intermediate stage.

\bibliographystyle{utphys}
\bibliography{massforisland}

\providecommand{\href}[2]{#2}\begingroup\raggedright\begin{thebibliography}{100}

\bibitem{Penington:2019npb}
G.~Penington, ``{Entanglement Wedge Reconstruction and the Information
  Paradox},'' \href{http://dx.doi.org/10.1007/JHEP09(2020)002}{{\em JHEP}
  {\bfseries 09} (2020) 002}, \href{http://arxiv.org/abs/1905.08255}{{\ttfamily
  arXiv:1905.08255 [hep-th]}}.

\bibitem{Almheiri:2019hni}
A.~Almheiri, R.~Mahajan, J.~Maldacena, and Y.~Zhao, ``{The Page curve of
  Hawking radiation from semiclassical geometry},''
  \href{http://dx.doi.org/10.1007/JHEP03(2020)149}{{\em JHEP} {\bfseries 03}
  (2020) 149}, \href{http://arxiv.org/abs/1908.10996}{{\ttfamily
  arXiv:1908.10996 [hep-th]}}.

\bibitem{Almheiri:2019psy}
A.~Almheiri, R.~Mahajan, and J.~E. Santos, ``{Entanglement islands in higher
  dimensions},'' \href{http://dx.doi.org/10.21468/SciPostPhys.9.1.001}{{\em
  SciPost Phys.} {\bfseries 9} no.~1, (2020) 001},
  \href{http://arxiv.org/abs/1911.09666}{{\ttfamily arXiv:1911.09666
  [hep-th]}}.

\bibitem{Almheiri:2019qdq}
A.~Almheiri, T.~Hartman, J.~Maldacena, E.~Shaghoulian, and A.~Tajdini,
  ``{Replica Wormholes and the Entropy of Hawking Radiation},''
  \href{http://dx.doi.org/10.1007/JHEP05(2020)013}{{\em JHEP} {\bfseries 05}
  (2020) 013}, \href{http://arxiv.org/abs/1911.12333}{{\ttfamily
  arXiv:1911.12333 [hep-th]}}.

\bibitem{Almheiri:2019psf}
A.~Almheiri, N.~Engelhardt, D.~Marolf, and H.~Maxfield, ``{The entropy of bulk
  quantum fields and the entanglement wedge of an evaporating black hole},''
  \href{http://dx.doi.org/10.1007/JHEP12(2019)063}{{\em JHEP} {\bfseries 12}
  (2019) 063}, \href{http://arxiv.org/abs/1905.08762}{{\ttfamily
  arXiv:1905.08762 [hep-th]}}.

\bibitem{Penington:2019kki}
G.~Penington, S.~H. Shenker, D.~Stanford, and Z.~Yang, ``{Replica wormholes and
  the black hole interior},'' \href{http://arxiv.org/abs/1911.11977}{{\ttfamily
  arXiv:1911.11977 [hep-th]}}.

\bibitem{Karch:2000ct}
A.~Karch and L.~Randall, ``{Locally localized gravity},''
  \href{http://dx.doi.org/10.1088/1126-6708/2001/05/008}{{\em JHEP} {\bfseries
  05} (2001) 008}, \href{http://arxiv.org/abs/hep-th/0011156}{{\ttfamily
  arXiv:hep-th/0011156}}.

\bibitem{Karch:2000gx}
A.~Karch and L.~Randall, ``{Open and closed string interpretation of SUSY CFT's
  on branes with boundaries},''
  \href{http://dx.doi.org/10.1088/1126-6708/2001/06/063}{{\em JHEP} {\bfseries
  06} (2001) 063}, \href{http://arxiv.org/abs/hep-th/0105132}{{\ttfamily
  arXiv:hep-th/0105132}}.

\bibitem{Rozali:2019day}
M.~Rozali, J.~Sully, M.~Van~Raamsdonk, C.~Waddell, and D.~Wakeham,
  ``{Information radiation in BCFT models of black holes},''
  \href{http://dx.doi.org/10.1007/JHEP05(2020)004}{{\em JHEP} {\bfseries 05}
  (2020) 004}, \href{http://arxiv.org/abs/1910.12836}{{\ttfamily
  arXiv:1910.12836 [hep-th]}}.

\bibitem{Liu:2020gnp}
H.~Liu and S.~Vardhan, ``{A dynamical mechanism for the Page curve from quantum
  chaos},'' \href{http://dx.doi.org/10.1007/JHEP03(2021)088}{{\em JHEP}
  {\bfseries 03} (2021) 088}, \href{http://arxiv.org/abs/2002.05734}{{\ttfamily
  arXiv:2002.05734 [hep-th]}}.

\bibitem{Sully:2020pza}
J.~Sully, M.~V. Raamsdonk, and D.~Wakeham, ``{BCFT entanglement entropy at
  large central charge and the black hole interior},''
  \href{http://dx.doi.org/10.1007/JHEP03(2021)167}{{\em JHEP} {\bfseries 03}
  (2021) 167}, \href{http://arxiv.org/abs/2004.13088}{{\ttfamily
  arXiv:2004.13088 [hep-th]}}.

\bibitem{Emparan:2020znc}
R.~Emparan, A.~M. Frassino, and B.~Way, ``{Quantum BTZ black hole},''
  \href{http://dx.doi.org/10.1007/JHEP11(2020)137}{{\em JHEP} {\bfseries 11}
  (2020) 137}, \href{http://arxiv.org/abs/2007.15999}{{\ttfamily
  arXiv:2007.15999 [hep-th]}}.

\bibitem{Liu:2020jsv}
H.~Liu and S.~Vardhan, ``{Entanglement entropies of equilibrated pure states in
  quantum many-body systems and gravity},''
  \href{http://dx.doi.org/10.1103/PRXQuantum.2.010344}{{\em P. R. X. Quantum.}
  {\bfseries 2} (2021) 010344},
  \href{http://arxiv.org/abs/2008.01089}{{\ttfamily arXiv:2008.01089
  [hep-th]}}.

\bibitem{Ling:2020laa}
Y.~Ling, Y.~Liu, and Z.-Y. Xian, ``{Island in Charged Black Holes},''
  \href{http://dx.doi.org/10.1007/JHEP03(2021)251}{{\em JHEP} {\bfseries 03}
  (2021) 251}, \href{http://arxiv.org/abs/2010.00037}{{\ttfamily
  arXiv:2010.00037 [hep-th]}}.

\bibitem{KumarBasak:2020ams}
J.~Kumar~Basak, D.~Basu, V.~Malvimat, H.~Parihar, and G.~Sengupta, ``{Islands
  for Entanglement Negativity},''
  \href{http://arxiv.org/abs/2012.03983}{{\ttfamily arXiv:2012.03983
  [hep-th]}}.

\bibitem{Caceres:2020jcn}
E.~Caceres, A.~Kundu, A.~K. Patra, and S.~Shashi, ``{Warped Information and
  Entanglement Islands in AdS/WCFT},''
  \href{http://dx.doi.org/10.1007/JHEP07(2021)004}{{\em JHEP} {\bfseries 07}
  (2021) 004}, \href{http://arxiv.org/abs/2012.05425}{{\ttfamily
  arXiv:2012.05425 [hep-th]}}.

\bibitem{Caceres:2021fuw}
E.~Caceres, A.~Kundu, A.~K. Patra, and S.~Shashi, ``{Page Curves and Bath
  Deformations},'' \href{http://arxiv.org/abs/2107.00022}{{\ttfamily
  arXiv:2107.00022 [hep-th]}}.

\bibitem{Deng:2020ent}
F.~Deng, J.~Chu, and Y.~Zhou, ``{Defect extremal surface as the holographic
  counterpart of Island formula},''
  \href{http://dx.doi.org/10.1007/JHEP03(2021)008}{{\em JHEP} {\bfseries 03}
  (2021) 008}, \href{http://arxiv.org/abs/2012.07612}{{\ttfamily
  arXiv:2012.07612 [hep-th]}}.

\bibitem{Karlsson:2021vlh}
A.~Karlsson, ``{Concerns about the replica wormhole derivation of the island
  conjecture},'' \href{http://arxiv.org/abs/2101.05879}{{\ttfamily
  arXiv:2101.05879 [hep-th]}}.

\bibitem{Miao:2021ual}
R.-X. Miao, ``{Codimension-n holography for cones},''
  \href{http://dx.doi.org/10.1103/PhysRevD.104.086031}{{\em Phys. Rev. D}
  {\bfseries 104} no.~8, (2021) 086031},
  \href{http://arxiv.org/abs/2101.10031}{{\ttfamily arXiv:2101.10031
  [hep-th]}}.

\bibitem{Bachas:2021fqo}
C.~Bachas and V.~Papadopoulos, ``{Phases of Holographic Interfaces},''
  \href{http://dx.doi.org/10.1007/JHEP04(2021)262}{{\em JHEP} {\bfseries 04}
  (2021) 262}, \href{http://arxiv.org/abs/2101.12529}{{\ttfamily
  arXiv:2101.12529 [hep-th]}}.

\bibitem{May:2021zyu}
A.~May and D.~Wakeham, ``{Quantum tasks require islands on the brane},''
  \href{http://dx.doi.org/10.1088/1361-6382/ac025d}{{\em Class. Quant. Grav.}
  {\bfseries 38} no.~14, (2021) 144001},
  \href{http://arxiv.org/abs/2102.01810}{{\ttfamily arXiv:2102.01810
  [hep-th]}}.

\bibitem{Kawabata:2021hac}
K.~Kawabata, T.~Nishioka, Y.~Okuyama, and K.~Watanabe, ``{Probing Hawking
  radiation through capacity of entanglement},''
  \href{http://dx.doi.org/10.1007/JHEP05(2021)062}{{\em JHEP} {\bfseries 05}
  (2021) 062}, \href{http://arxiv.org/abs/2102.02425}{{\ttfamily
  arXiv:2102.02425 [hep-th]}}.

\bibitem{Bhattacharya:2021jrn}
A.~Bhattacharya, A.~Bhattacharyya, P.~Nandy, and A.~K. Patra, ``{Islands and
  complexity of eternal black hole and radiation subsystems for a doubly
  holographic model},'' \href{http://dx.doi.org/10.1007/JHEP05(2021)135}{{\em
  JHEP} {\bfseries 05} (2021) 135},
  \href{http://arxiv.org/abs/2103.15852}{{\ttfamily arXiv:2103.15852
  [hep-th]}}.

\bibitem{Kim:2021gzd}
W.~Kim and M.~Nam, ``{Entanglement entropy of asymptotically flat non-extremal
  and extremal black holes with an island},''
  \href{http://dx.doi.org/10.1140/epjc/s10052-021-09680-x}{{\em Eur. Phys. J.
  C} {\bfseries 81} no.~10, (2021) 869},
  \href{http://arxiv.org/abs/2103.16163}{{\ttfamily arXiv:2103.16163
  [hep-th]}}.

\bibitem{Aalsma:2021bit}
L.~Aalsma and W.~Sybesma, ``{The Price of Curiosity: Information Recovery in de
  Sitter Space},'' \href{http://dx.doi.org/10.1007/JHEP05(2021)291}{{\em JHEP}
  {\bfseries 05} (2021) 291}, \href{http://arxiv.org/abs/2104.00006}{{\ttfamily
  arXiv:2104.00006 [hep-th]}}.

\bibitem{Neuenfeld:2021wbl}
D.~Neuenfeld, ``{The Dictionary for Double Holography and Graviton Masses in d
  Dimensions},'' \href{http://arxiv.org/abs/2104.02801}{{\ttfamily
  arXiv:2104.02801 [hep-th]}}.

\bibitem{Geng:2021iyq}
H.~Geng, S.~L\"ust, R.~K. Mishra, and D.~Wakeham, ``{Holographic BCFTs and
  Communicating Black Holes},''
  \href{http://dx.doi.org/10.1007/JHEP08(2021)003}{{\em jhep} {\bfseries 08}
  (2021) 003}, \href{http://arxiv.org/abs/2104.07039}{{\ttfamily
  arXiv:2104.07039 [hep-th]}}.

\bibitem{Balasubramanian:2021wgd}
V.~Balasubramanian, A.~Kar, and T.~Ugajin, ``{Entanglement between two
  gravitating universes},'' \href{http://arxiv.org/abs/2104.13383}{{\ttfamily
  arXiv:2104.13383 [hep-th]}}.

\bibitem{Uhlemann:2021nhu}
C.~F. Uhlemann, ``{Islands and Page curves in 4d from Type IIB},''
  \href{http://dx.doi.org/10.1007/JHEP08(2021)104}{{\em JHEP} {\bfseries 08}
  (2021) 104}, \href{http://arxiv.org/abs/2105.00008}{{\ttfamily
  arXiv:2105.00008 [hep-th]}}.

\bibitem{Neuenfeld:2021bsb}
D.~Neuenfeld, ``{Double Holography as a Model for Black Hole
  Complementarity},'' \href{http://arxiv.org/abs/2105.01130}{{\ttfamily
  arXiv:2105.01130 [hep-th]}}.

\bibitem{Kawabata:2021vyo}
K.~Kawabata, T.~Nishioka, Y.~Okuyama, and K.~Watanabe, ``{Replica wormholes and
  capacity of entanglement},''
  \href{http://dx.doi.org/10.1007/JHEP10(2021)227}{{\em JHEP} {\bfseries 10}
  (2021) 227}, \href{http://arxiv.org/abs/2105.08396}{{\ttfamily
  arXiv:2105.08396 [hep-th]}}.

\bibitem{Chu:2021gdb}
J.~Chu, F.~Deng, and Y.~Zhou, ``{Page curve from defect extremal surface and
  island in higher dimensions},''
  \href{http://dx.doi.org/10.1007/JHEP10(2021)149}{{\em JHEP} {\bfseries 10}
  (2021) 149}, \href{http://arxiv.org/abs/2105.09106}{{\ttfamily
  arXiv:2105.09106 [hep-th]}}.

\bibitem{Kruthoff:2021vgv}
J.~Kruthoff, R.~Mahajan, and C.~Murdia, ``{Free fermion entanglement with a
  semitransparent interface: the effect of graybody factors on entanglement
  islands},'' \href{http://dx.doi.org/10.21468/SciPostPhys.11.3.063}{{\em
  SciPost Phys.} {\bfseries 11} (2021) 063},
  \href{http://arxiv.org/abs/2106.10287}{{\ttfamily arXiv:2106.10287
  [hep-th]}}.

\bibitem{Akal:2021foz}
I.~Akal, Y.~Kusuki, N.~Shiba, T.~Takayanagi, and Z.~Wei, ``{Holographic moving
  mirrors},'' \href{http://dx.doi.org/10.1088/1361-6382/ac2c1b}{{\em Class.
  Quant. Grav.} {\bfseries 38} no.~22, (2021) 224001},
  \href{http://arxiv.org/abs/2106.11179}{{\ttfamily arXiv:2106.11179
  [hep-th]}}.

\bibitem{KumarBasak:2021rrx}
J.~Kumar~Basak, D.~Basu, V.~Malvimat, H.~Parihar, and G.~Sengupta, ``{Page
  Curve for Entanglement Negativity through Geometric Evaporation},''
  \href{http://arxiv.org/abs/2106.12593}{{\ttfamily arXiv:2106.12593
  [hep-th]}}.

\bibitem{Lu:2021gmv}
Y.~Lu and J.~Lin, ``{Islands in Kaluza-Klein black holes},''
  \href{http://arxiv.org/abs/2106.07845}{{\ttfamily arXiv:2106.07845
  [hep-th]}}.

\bibitem{Omiya:2021olc}
H.~Omiya and Z.~Wei, ``{Causal Structures and Nonlocality in Double
  Holography},'' \href{http://arxiv.org/abs/2107.01219}{{\ttfamily
  arXiv:2107.01219 [hep-th]}}.

\bibitem{Manu:2020tty}
A.~Manu, K.~Narayan, and P.~Paul, ``{Cosmological singularities, entanglement
  and quantum extremal surfaces},''
  \href{http://dx.doi.org/10.1007/JHEP04(2021)200}{{\em JHEP} {\bfseries 04}
  (2021) 200}, \href{http://arxiv.org/abs/2012.07351}{{\ttfamily
  arXiv:2012.07351 [hep-th]}}.

\bibitem{Geng:2020qvw}
H.~Geng and A.~Karch, ``{Massive islands},''
  \href{http://dx.doi.org/10.1007/JHEP09(2020)121}{{\em JHEP} {\bfseries 09}
  (2020) 121}, \href{http://arxiv.org/abs/2006.02438}{{\ttfamily
  arXiv:2006.02438 [hep-th]}}.

\bibitem{Krishnan:2020fer}
C.~Krishnan, ``{Critical Islands},''
  \href{http://dx.doi.org/10.1007/JHEP01(2021)179}{{\em JHEP} {\bfseries 01}
  (2021) 179}, \href{http://arxiv.org/abs/2007.06551}{{\ttfamily
  arXiv:2007.06551 [hep-th]}}.

\bibitem{Almheiri:2020cfm}
A.~Almheiri, T.~Hartman, J.~Maldacena, E.~Shaghoulian, and A.~Tajdini, ``{The
  entropy of Hawking radiation},''
  \href{http://dx.doi.org/10.1103/RevModPhys.93.035002}{{\em Rev. Mod. Phys.}
  {\bfseries 93} no.~3, (2021) 035002},
  \href{http://arxiv.org/abs/2006.06872}{{\ttfamily arXiv:2006.06872
  [hep-th]}}.

\bibitem{Geng:2020fxl}
H.~Geng, A.~Karch, C.~Perez-Pardavila, S.~Raju, L.~Randall, M.~Riojas, and
  S.~Shashi, ``{Information Transfer with a Gravitating Bath},''
  \href{http://dx.doi.org/10.21468/SciPostPhys.10.5.103}{{\em SciPost Phys.}
  {\bfseries 10} no.~5, (2021) 103},
  \href{http://arxiv.org/abs/2012.04671}{{\ttfamily arXiv:2012.04671
  [hep-th]}}.

\bibitem{Laddha:2020kvp}
A.~Laddha, S.~G. Prabhu, S.~Raju, and P.~Shrivastava, ``{The Holographic Nature
  of Null Infinity},''
  \href{http://dx.doi.org/10.21468/SciPostPhys.10.2.041}{{\em SciPost Phys.}
  {\bfseries 10} no.~2, (2021) 041},
  \href{http://arxiv.org/abs/2002.02448}{{\ttfamily arXiv:2002.02448
  [hep-th]}}.

\bibitem{Ghosh:2021axl}
K.~Ghosh and C.~Krishnan, ``{Dirichlet baths and the not-so-fine-grained Page
  curve},'' \href{http://dx.doi.org/10.1007/JHEP08(2021)119}{{\em JHEP}
  {\bfseries 08} (2021) 119}, \href{http://arxiv.org/abs/2103.17253}{{\ttfamily
  arXiv:2103.17253 [hep-th]}}.

\bibitem{Geng:2021wcq}
H.~Geng, Y.~Nomura, and H.-Y. Sun, ``{Information paradox and its resolution in
  de Sitter holography},''
  \href{http://dx.doi.org/10.1103/PhysRevD.103.126004}{{\em Phys. Rev. D}
  {\bfseries 103} no.~12, (2021) 126004},
  \href{http://arxiv.org/abs/2103.07477}{{\ttfamily arXiv:2103.07477
  [hep-th]}}.

\bibitem{Langhoff:2020jqa}
K.~Langhoff and Y.~Nomura, ``{Ensemble from Coarse Graining: Reconstructing the
  Interior of an Evaporating Black Hole},''
  \href{http://dx.doi.org/10.1103/PhysRevD.102.086021}{{\em Phys. Rev. D}
  {\bfseries 102} no.~8, (2020) 086021},
  \href{http://arxiv.org/abs/2008.04202}{{\ttfamily arXiv:2008.04202
  [hep-th]}}.

\bibitem{Maldacena:1997re}
J.~M. Maldacena, ``{The Large N limit of superconformal field theories and
  supergravity},'' \href{http://dx.doi.org/10.1023/A:1026654312961}{{\em Adv.
  Theor. Math. Phys.} {\bfseries 2} (1998) 231--252},
  \href{http://arxiv.org/abs/hep-th/9711200}{{\ttfamily arXiv:hep-th/9711200}}.

\bibitem{Gubser:1998bc}
S.~S. Gubser, I.~R. Klebanov, and A.~M. Polyakov, ``{Gauge theory correlators
  from noncritical string theory},''
  \href{http://dx.doi.org/10.1016/S0370-2693(98)00377-3}{{\em Phys. Lett. B}
  {\bfseries 428} (1998) 105--114},
  \href{http://arxiv.org/abs/hep-th/9802109}{{\ttfamily arXiv:hep-th/9802109}}.

\bibitem{Witten:1998qj}
E.~Witten, ``{Anti-de Sitter space and holography},''
  \href{http://dx.doi.org/10.4310/ATMP.1998.v2.n2.a2}{{\em Adv. Theor. Math.
  Phys.} {\bfseries 2} (1998) 253--291},
  \href{http://arxiv.org/abs/hep-th/9802150}{{\ttfamily arXiv:hep-th/9802150}}.

\bibitem{Chowdhury:2020hse}
C.~Chowdhury, O.~Papadoulaki, and S.~Raju, ``{A physical protocol for observers
  near the boundary to obtain bulk information in quantum gravity},''
  \href{http://dx.doi.org/10.21468/SciPostPhys.10.5.106}{{\em SciPost Phys.}
  {\bfseries 10} no.~5, (2021) 106},
  \href{http://arxiv.org/abs/2008.01740}{{\ttfamily arXiv:2008.01740
  [hep-th]}}.

\bibitem{Raju:2020smc}
S.~Raju, ``{Lessons from the information paradox},''
  \href{http://dx.doi.org/10.1016/j.physrep.2021.10.001}{{\em Phys. Rept.}
  {\bfseries 943} (2022) 2187},
  \href{http://arxiv.org/abs/2012.05770}{{\ttfamily arXiv:2012.05770
  [hep-th]}}.

\bibitem{Aharony:2003qf}
O.~Aharony, O.~DeWolfe, D.~Z. Freedman, and A.~Karch, ``{Defect conformal field
  theory and locally localized gravity},''
  \href{http://dx.doi.org/10.1088/1126-6708/2003/07/030}{{\em JHEP} {\bfseries
  07} (2003) 030}, \href{http://arxiv.org/abs/hep-th/0303249}{{\ttfamily
  arXiv:hep-th/0303249}}.

\bibitem{Ryu:2006bv}
S.~Ryu and T.~Takayanagi, ``{Holographic derivation of entanglement entropy
  from AdS/CFT},'' \href{http://dx.doi.org/10.1103/PhysRevLett.96.181602}{{\em
  Phys. Rev. Lett.} {\bfseries 96} (2006) 181602},
  \href{http://arxiv.org/abs/hep-th/0603001}{{\ttfamily arXiv:hep-th/0603001}}.

\bibitem{Ryu:2006ef}
S.~Ryu and T.~Takayanagi, ``{Aspects of Holographic Entanglement Entropy},''
  \href{http://dx.doi.org/10.1088/1126-6708/2006/08/045}{{\em JHEP} {\bfseries
  08} (2006) 045}, \href{http://arxiv.org/abs/hep-th/0605073}{{\ttfamily
  arXiv:hep-th/0605073}}.

\bibitem{Hubeny:2007xt}
V.~E. Hubeny, M.~Rangamani, and T.~Takayanagi, ``{A Covariant holographic
  entanglement entropy proposal},''
  \href{http://dx.doi.org/10.1088/1126-6708/2007/07/062}{{\em JHEP} {\bfseries
  07} (2007) 062}, \href{http://arxiv.org/abs/0705.0016}{{\ttfamily
  arXiv:0705.0016 [hep-th]}}.

\bibitem{Porrati:2003sa}
M.~Porrati, ``{Higgs phenomenon for the graviton in ADS space},''
  \href{http://dx.doi.org/10.1142/S0217732303011745}{{\em Mod. Phys. Lett. A}
  {\bfseries 18} (2003) 1793--1802},
  \href{http://arxiv.org/abs/hep-th/0306253}{{\ttfamily arXiv:hep-th/0306253}}.

\bibitem{Porrati:2001gx}
M.~Porrati, ``{Mass and gauge invariance 4. Holography for the Karch-Randall
  model},'' \href{http://dx.doi.org/10.1103/PhysRevD.65.044015}{{\em Phys. Rev.
  D} {\bfseries 65} (2002) 044015},
  \href{http://arxiv.org/abs/hep-th/0109017}{{\ttfamily arXiv:hep-th/0109017}}.

\bibitem{Porrati:2002dt}
M.~Porrati and A.~Starinets, ``{On the graviton selfenergy in AdS(4)},''
  \href{http://dx.doi.org/10.1016/S0370-2693(02)01490-9}{{\em Phys. Lett. B}
  {\bfseries 532} (2002) 48--54},
  \href{http://arxiv.org/abs/hep-th/0201261}{{\ttfamily arXiv:hep-th/0201261}}.

\bibitem{Headrick:2014cta}
M.~Headrick, V.~E. Hubeny, A.~Lawrence, and M.~Rangamani, ``{Causality \&
  holographic entanglement entropy},''
  \href{http://dx.doi.org/10.1007/JHEP12(2014)162}{{\em JHEP} {\bfseries 12}
  (2014) 162}, \href{http://arxiv.org/abs/1408.6300}{{\ttfamily arXiv:1408.6300
  [hep-th]}}.

\bibitem{Hamilton:2005ju}
A.~Hamilton, D.~N. Kabat, G.~Lifschytz, and D.~A. Lowe, ``{Local bulk operators
  in AdS/CFT: A Boundary view of horizons and locality},''
  \href{http://dx.doi.org/10.1103/PhysRevD.73.086003}{{\em Phys. Rev. D}
  {\bfseries 73} (2006) 086003},
  \href{http://arxiv.org/abs/hep-th/0506118}{{\ttfamily arXiv:hep-th/0506118}}.

\bibitem{Faulkner:2017vdd}
T.~Faulkner and A.~Lewkowycz, ``{Bulk locality from modular flow},''
  \href{http://dx.doi.org/10.1007/JHEP07(2017)151}{{\em JHEP} {\bfseries 07}
  (2017) 151}, \href{http://arxiv.org/abs/1704.05464}{{\ttfamily
  arXiv:1704.05464 [hep-th]}}.

\bibitem{Engelhardt:2014gca}
N.~Engelhardt and A.~C. Wall, ``{Quantum Extremal Surfaces: Holographic
  Entanglement Entropy beyond the Classical Regime},''
  \href{http://dx.doi.org/10.1007/JHEP01(2015)073}{{\em JHEP} {\bfseries 01}
  (2015) 073}, \href{http://arxiv.org/abs/1408.3203}{{\ttfamily arXiv:1408.3203
  [hep-th]}}.

\bibitem{Faulkner:2013ana}
T.~Faulkner, A.~Lewkowycz, and J.~Maldacena, ``{Quantum corrections to
  holographic entanglement entropy},''
  \href{http://dx.doi.org/10.1007/JHEP11(2013)074}{{\em JHEP} {\bfseries 11}
  (2013) 074}, \href{http://arxiv.org/abs/1307.2892}{{\ttfamily arXiv:1307.2892
  [hep-th]}}.

\bibitem{Headrick:2013zda}
M.~Headrick, ``{General properties of holographic entanglement entropy},''
  \href{http://dx.doi.org/10.1007/JHEP03(2014)085}{{\em JHEP} {\bfseries 03}
  (2014) 085}, \href{http://arxiv.org/abs/1312.6717}{{\ttfamily arXiv:1312.6717
  [hep-th]}}.

\bibitem{Jafferis:2015del}
D.~L. Jafferis, A.~Lewkowycz, J.~Maldacena, and S.~J. Suh, ``{Relative entropy
  equals bulk relative entropy},''
  \href{http://dx.doi.org/10.1007/JHEP06(2016)004}{{\em JHEP} {\bfseries 06}
  (2016) 004}, \href{http://arxiv.org/abs/1512.06431}{{\ttfamily
  arXiv:1512.06431 [hep-th]}}.

\bibitem{Dong:2016eik}
X.~Dong, D.~Harlow, and A.~C. Wall, ``{Reconstruction of Bulk Operators within
  the Entanglement Wedge in Gauge-Gravity Duality},''
  \href{http://dx.doi.org/10.1103/PhysRevLett.117.021601}{{\em Phys. Rev.
  Lett.} {\bfseries 117} no.~2, (2016) 021601},
  \href{http://arxiv.org/abs/1601.05416}{{\ttfamily arXiv:1601.05416
  [hep-th]}}.

\bibitem{Haag:1992hx}
R.~Haag, {\em {Local quantum physics: Fields, particles, algebras}}.
\newblock Springer, 1992.

\bibitem{Casini:2013rba}
H.~Casini, M.~Huerta, and J.~A. Rosabal, ``{Remarks on entanglement entropy for
  gauge fields},'' \href{http://dx.doi.org/10.1103/PhysRevD.89.085012}{{\em
  Phys. Rev. D} {\bfseries 89} no.~8, (2014) 085012},
  \href{http://arxiv.org/abs/1312.1183}{{\ttfamily arXiv:1312.1183 [hep-th]}}.

\bibitem{Ghosh:2015iwa}
S.~Ghosh, R.~M. Soni, and S.~P. Trivedi, ``{On The Entanglement Entropy For
  Gauge Theories},'' \href{http://dx.doi.org/10.1007/JHEP09(2015)069}{{\em
  JHEP} {\bfseries 09} (2015) 069},
  \href{http://arxiv.org/abs/1501.02593}{{\ttfamily arXiv:1501.02593
  [hep-th]}}.

\bibitem{dewitt1960quantization}
B.~S. DeWitt, ``The quantization of geometry,'' in {\em Gravitation: an
  introduction to current research}, L.~Witten, ed.
\newblock John Wiley \& Sons, 1963.

\bibitem{kuchar1991problem}
K.~Kuchar, ``The problem of time in canonical quantization,'' in {\em
  Conceptual Problems of Quantum Gravity}, {A. Ashtekar} and {J. Stachel}, eds.
\newblock Birkhauser, 1991.

\bibitem{Giddings:2005id}
S.~B. Giddings, D.~Marolf, and J.~B. Hartle, ``{Observables in effective
  gravity},'' \href{http://dx.doi.org/10.1103/PhysRevD.74.064018}{{\em Phys.
  Rev. D} {\bfseries 74} (2006) 064018},
  \href{http://arxiv.org/abs/hep-th/0512200}{{\ttfamily arXiv:hep-th/0512200}}.

\bibitem{Jacobson:2012gh}
T.~Jacobson, ``{Boundary unitarity and the black hole information paradox},''
  \href{http://dx.doi.org/10.1142/S0218271813420029}{{\em Int. J. Mod. Phys. D}
  {\bfseries 22} (2013) 1342002},
  \href{http://arxiv.org/abs/1212.6944}{{\ttfamily arXiv:1212.6944 [hep-th]}}.

\bibitem{Jacobson:2019gnm}
T.~Jacobson and P.~Nguyen, ``{Diffeomorphism invariance and the black hole
  information paradox},''
  \href{http://dx.doi.org/10.1103/PhysRevD.100.046002}{{\em Phys. Rev. D}
  {\bfseries 100} no.~4, (2019) 046002},
  \href{http://arxiv.org/abs/1904.04434}{{\ttfamily arXiv:1904.04434 [gr-qc]}}.

\bibitem{Banks:1998dd}
T.~Banks, M.~R. Douglas, G.~T. Horowitz, and E.~J. Martinec, ``{AdS dynamics
  from conformal field theory},''
  \href{http://arxiv.org/abs/hep-th/9808016}{{\ttfamily arXiv:hep-th/9808016}}.

\bibitem{deHaro:2000vlm}
S.~de~Haro, S.~N. Solodukhin, and K.~Skenderis, ``{Holographic reconstruction
  of space-time and renormalization in the AdS / CFT correspondence},''
  \href{http://dx.doi.org/10.1007/s002200100381}{{\em Commun. Math. Phys.}
  {\bfseries 217} (2001) 595--622},
  \href{http://arxiv.org/abs/hep-th/0002230}{{\ttfamily arXiv:hep-th/0002230}}.

\bibitem{Almheiri:2014lwa}
A.~Almheiri, X.~Dong, and D.~Harlow, ``{Bulk Locality and Quantum Error
  Correction in AdS/CFT},''
  \href{http://dx.doi.org/10.1007/JHEP04(2015)163}{{\em JHEP} {\bfseries 04}
  (2015) 163}, \href{http://arxiv.org/abs/1411.7041}{{\ttfamily arXiv:1411.7041
  [hep-th]}}.

\bibitem{Papadodimas:2013jku}
K.~Papadodimas and S.~Raju, ``{State-Dependent Bulk-Boundary Maps and Black
  Hole Complementarity},''
  \href{http://dx.doi.org/10.1103/PhysRevD.89.086010}{{\em Phys. Rev. D}
  {\bfseries 89} no.~8, (2014) 086010},
  \href{http://arxiv.org/abs/1310.6335}{{\ttfamily arXiv:1310.6335 [hep-th]}}.

\bibitem{Almheiri:2019yqk}
A.~Almheiri, R.~Mahajan, and J.~Maldacena, ``{Islands outside the horizon},''
  \href{http://arxiv.org/abs/1910.11077}{{\ttfamily arXiv:1910.11077
  [hep-th]}}.

\bibitem{Bousso:2020kmy}
R.~Bousso and E.~Wildenhain, ``{Gravity/ensemble duality},''
  \href{http://dx.doi.org/10.1103/PhysRevD.102.066005}{{\em Phys. Rev. D}
  {\bfseries 102} no.~6, (2020) 066005},
  \href{http://arxiv.org/abs/2006.16289}{{\ttfamily arXiv:2006.16289
  [hep-th]}}.

\bibitem{ElShowk:2011ag}
S.~El-Showk and K.~Papadodimas, ``{Emergent Spacetime and Holographic CFTs},''
  \href{http://dx.doi.org/10.1007/JHEP10(2012)106}{{\em JHEP} {\bfseries 10}
  (2012) 106}, \href{http://arxiv.org/abs/1101.4163}{{\ttfamily arXiv:1101.4163
  [hep-th]}}.

\bibitem{Papadodimas:2015jra}
K.~Papadodimas and S.~Raju, ``{Remarks on the necessity and implications of
  state-dependence in the black hole interior},''
  \href{http://dx.doi.org/10.1103/PhysRevD.93.084049}{{\em Phys. Rev. D}
  {\bfseries 93} no.~8, (2016) 084049},
  \href{http://arxiv.org/abs/1503.08825}{{\ttfamily arXiv:1503.08825
  [hep-th]}}.

\bibitem{Papadodimas:2017qit}
K.~Papadodimas, ``{A class of non-equilibrium states and the black hole
  interior},'' \href{http://arxiv.org/abs/1708.06328}{{\ttfamily
  arXiv:1708.06328 [hep-th]}}.

\bibitem{Hollowood:2021nlo}
T.~J. Hollowood, S.~P. Kumar, A.~Legramandi, and N.~Talwar, ``{Islands in the
  stream of Hawking radiation},''
  \href{http://dx.doi.org/10.1007/JHEP11(2021)067}{{\em JHEP} {\bfseries 11}
  (2021) 067}, \href{http://arxiv.org/abs/2104.00052}{{\ttfamily
  arXiv:2104.00052 [hep-th]}}.

\bibitem{Wang:2021mqq}
X.~Wang, R.~Li, and J.~Wang, ``{Page curves for a family of exactly solvable
  evaporating black holes},''
  \href{http://dx.doi.org/10.1103/PhysRevD.103.126026}{{\em Phys. Rev. D}
  {\bfseries 103} no.~12, (2021) 126026},
  \href{http://arxiv.org/abs/2104.00224}{{\ttfamily arXiv:2104.00224
  [hep-th]}}.

\bibitem{Wang:2021woy}
X.~Wang, R.~Li, and J.~Wang, ``{Islands and Page curves of Reissner-Nordstr\"om
  black holes},'' \href{http://dx.doi.org/10.1007/JHEP04(2021)103}{{\em JHEP}
  {\bfseries 04} (2021) 103}, \href{http://arxiv.org/abs/2101.06867}{{\ttfamily
  arXiv:2101.06867 [hep-th]}}.

\bibitem{Takayanagi:2011zk}
T.~Takayanagi, ``{Holographic Dual of BCFT},''
  \href{http://dx.doi.org/10.1103/PhysRevLett.107.101602}{{\em Phys. Rev.
  Lett.} {\bfseries 107} (2011) 101602},
  \href{http://arxiv.org/abs/1105.5165}{{\ttfamily arXiv:1105.5165 [hep-th]}}.

\bibitem{Gaiotto:2008sa}
D.~Gaiotto and E.~Witten, ``{Supersymmetric Boundary Conditions in N=4 Super
  Yang-Mills Theory},'' \href{http://dx.doi.org/10.1007/s10955-009-9687-3}{{\em
  J. Statist. Phys.} {\bfseries 135} (2009) 789--855},
  \href{http://arxiv.org/abs/0804.2902}{{\ttfamily arXiv:0804.2902 [hep-th]}}.

\bibitem{Hartman:2013qma}
T.~Hartman and J.~Maldacena, ``{Time Evolution of Entanglement Entropy from
  Black Hole Interiors},''
  \href{http://dx.doi.org/10.1007/JHEP05(2013)014}{{\em JHEP} {\bfseries 05}
  (2013) 014}, \href{http://arxiv.org/abs/1303.1080}{{\ttfamily arXiv:1303.1080
  [hep-th]}}.

\bibitem{Aharony:2006hz}
O.~Aharony, A.~B. Clark, and A.~Karch, ``{The CFT/AdS correspondence, massive
  gravitons and a connectivity index conjecture},''
  \href{http://dx.doi.org/10.1103/PhysRevD.74.086006}{{\em Phys. Rev. D}
  {\bfseries 74} (2006) 086006},
  \href{http://arxiv.org/abs/hep-th/0608089}{{\ttfamily arXiv:hep-th/0608089}}.

\bibitem{vanDam:1970vg}
H.~van Dam and M.~J.~G. Veltman, ``{Massive and massless Yang-Mills and
  gravitational fields},''
  \href{http://dx.doi.org/10.1016/0550-3213(70)90416-5}{{\em Nucl. Phys. B}
  {\bfseries 22} (1970) 397--411}.

\bibitem{Zakharov:1970cc}
V.~I. Zakharov, ``{Linearized gravitation theory and the graviton mass},'' {\em
  JETP Lett.} {\bfseries 12} (1970) 312.

\bibitem{Arnowitt:1962hi}
R.~L. Arnowitt, S.~Deser, and C.~W. Misner, ``{The Dynamics of general
  relativity},'' \href{http://dx.doi.org/10.1007/s10714-008-0661-1}{{\em Gen.
  Rel. Grav.} {\bfseries 40} (2008) 1997--2027},
  \href{http://arxiv.org/abs/gr-qc/0405109}{{\ttfamily arXiv:gr-qc/0405109}}.

\bibitem{Kuchar:1970mu}
K.~Kuchar, ``{Ground state functional of the linearized gravitational field},''
  \href{http://dx.doi.org/10.1063/1.1665133}{{\em J. Math. Phys.} {\bfseries
  11} (1970) 3322--3334}.

\bibitem{Corvino:2003sp}
J.~Corvino and R.~M. Schoen, ``{On the asymptotics for the vacuum Einstein
  constraint equations},'' {\em J. Diff. Geom.} {\bfseries 73} no.~2, (2006)
  185--217, \href{http://arxiv.org/abs/gr-qc/0301071}{{\ttfamily
  arXiv:gr-qc/0301071}}.

\bibitem{Hinterbichler:2011tt}
K.~Hinterbichler, ``{Theoretical Aspects of Massive Gravity},''
  \href{http://dx.doi.org/10.1103/RevModPhys.84.671}{{\em Rev. Mod. Phys.}
  {\bfseries 84} (2012) 671--710},
  \href{http://arxiv.org/abs/1105.3735}{{\ttfamily arXiv:1105.3735 [hep-th]}}.

\bibitem{Goldhaber:2008xy}
A.~S. Goldhaber and M.~M. Nieto, ``{Photon and Graviton Mass Limits},''
  \href{http://dx.doi.org/10.1103/RevModPhys.82.939}{{\em Rev. Mod. Phys.}
  {\bfseries 82} (2010) 939--979},
  \href{http://arxiv.org/abs/0809.1003}{{\ttfamily arXiv:0809.1003 [hep-ph]}}.

\bibitem{DeWitt:1967yk}
B.~S. DeWitt, ``{Quantum Theory of Gravity. 1. The Canonical Theory},''
  \href{http://dx.doi.org/10.1103/PhysRev.160.1113}{{\em Phys. Rev.} {\bfseries
  160} (1967) 1113--1148}.

\bibitem{Chowdhury:2021nxw}
C.~Chowdhury, V.~Godet, O.~Papadoulaki, and S.~Raju, ``{Holography from the
  Wheeler-DeWitt equation},'' \href{http://arxiv.org/abs/2107.14802}{{\ttfamily
  arXiv:2107.14802 [hep-th]}}.

\bibitem{Hawking:1995fd}
S.~W. Hawking and G.~T. Horowitz, ``{The Gravitational Hamiltonian, action,
  entropy and surface terms},''
  \href{http://dx.doi.org/10.1088/0264-9381/13/6/017}{{\em Class. Quant. Grav.}
  {\bfseries 13} (1996) 1487--1498},
  \href{http://arxiv.org/abs/gr-qc/9501014}{{\ttfamily arXiv:gr-qc/9501014}}.

\bibitem{Balasubramanian:1999re}
V.~Balasubramanian and P.~Kraus, ``{A Stress tensor for Anti-de Sitter
  gravity},'' \href{http://dx.doi.org/10.1007/s002200050764}{{\em Commun. Math.
  Phys.} {\bfseries 208} (1999) 413--428},
  \href{http://arxiv.org/abs/hep-th/9902121}{{\ttfamily arXiv:hep-th/9902121}}.

\bibitem{Hartman:2020khs}
T.~Hartman, Y.~Jiang, and E.~Shaghoulian, ``{Islands in cosmology},''
  \href{http://dx.doi.org/10.1007/JHEP11(2020)111}{{\em JHEP} {\bfseries 11}
  (2020) 111}, \href{http://arxiv.org/abs/2008.01022}{{\ttfamily
  arXiv:2008.01022 [hep-th]}}.

\bibitem{Balasubramanian:2020coy}
V.~Balasubramanian, A.~Kar, and T.~Ugajin, ``{Entanglement between two disjoint
  universes},'' \href{http://dx.doi.org/10.1007/JHEP02(2021)136}{{\em JHEP}
  {\bfseries 02} (2021) 136}, \href{http://arxiv.org/abs/2008.05274}{{\ttfamily
  arXiv:2008.05274 [hep-th]}}.

\bibitem{Anderson:2021vof}
L.~Anderson, O.~Parrikar, and R.~M. Soni, ``{Islands with gravitating baths:
  towards ER = EPR},'' \href{http://dx.doi.org/10.1007/JHEP10(2021)226}{{\em
  JHEP} {\bfseries 21} (2020) 226},
  \href{http://arxiv.org/abs/2103.14746}{{\ttfamily arXiv:2103.14746
  [hep-th]}}.

\bibitem{Fallows:2021sge}
S.~Fallows and S.~F. Ross, ``{Islands and mixed states in closed universes},''
  \href{http://dx.doi.org/10.1007/JHEP07(2021)022}{{\em JHEP} {\bfseries 07}
  (2021) 022}, \href{http://arxiv.org/abs/2103.14364}{{\ttfamily
  arXiv:2103.14364 [hep-th]}}.

\bibitem{Miyata:2021ncm}
A.~Miyata and T.~Ugajin, ``{Evaporation of black holes in flat space entangled
  with an auxiliary universe},''
  \href{http://arxiv.org/abs/2104.00183}{{\ttfamily arXiv:2104.00183
  [hep-th]}}.

\bibitem{Israel:1976ur}
W.~Israel, ``{Thermo field dynamics of black holes},''
  \href{http://dx.doi.org/10.1016/0375-9601(76)90178-X}{{\em Phys. Lett. A}
  {\bfseries 57} (1976) 107--110}.

\bibitem{Maldacena:2001kr}
J.~M. Maldacena, ``{Eternal black holes in anti-de Sitter},''
  \href{http://dx.doi.org/10.1088/1126-6708/2003/04/021}{{\em JHEP} {\bfseries
  04} (2003) 021}, \href{http://arxiv.org/abs/hep-th/0106112}{{\ttfamily
  arXiv:hep-th/0106112}}.

\bibitem{Donnelly:2016rvo}
W.~Donnelly and S.~B. Giddings, ``{Observables, gravitational dressing, and
  obstructions to locality and subsystems},''
  \href{http://dx.doi.org/10.1103/PhysRevD.94.104038}{{\em Phys. Rev. D}
  {\bfseries 94} no.~10, (2016) 104038},
  \href{http://arxiv.org/abs/1607.01025}{{\ttfamily arXiv:1607.01025
  [hep-th]}}.

\bibitem{Giddings:2019hjc}
S.~B. Giddings, ``{Gravitational dressing, soft charges, and perturbative
  gravitational splitting},''
  \href{http://dx.doi.org/10.1103/PhysRevD.100.126001}{{\em Phys. Rev. D}
  {\bfseries 100} no.~12, (2019) 126001},
  \href{http://arxiv.org/abs/1903.06160}{{\ttfamily arXiv:1903.06160
  [hep-th]}}.

\bibitem{Hartman:2020swn}
T.~Hartman, E.~Shaghoulian, and A.~Strominger, ``{Islands in Asymptotically
  Flat 2D Gravity},'' \href{http://dx.doi.org/10.1007/JHEP07(2020)022}{{\em
  JHEP} {\bfseries 07} (2020) 022},
  \href{http://arxiv.org/abs/2004.13857}{{\ttfamily arXiv:2004.13857
  [hep-th]}}.

\bibitem{Akal:2020twv}
I.~Akal, Y.~Kusuki, N.~Shiba, T.~Takayanagi, and Z.~Wei, ``{Entanglement
  Entropy in a Holographic Moving Mirror and the Page Curve},''
  \href{http://dx.doi.org/10.1103/PhysRevLett.126.061604}{{\em Phys. Rev.
  Lett.} {\bfseries 126} no.~6, (2021) 061604},
  \href{http://arxiv.org/abs/2011.12005}{{\ttfamily arXiv:2011.12005
  [hep-th]}}.

\bibitem{Gautason:2020tmk}
F.~F. Gautason, L.~Schneiderbauer, W.~Sybesma, and L.~Thorlacius, ``{Page Curve
  for an Evaporating Black Hole},''
  \href{http://dx.doi.org/10.1007/JHEP05(2020)091}{{\em JHEP} {\bfseries 05}
  (2020) 091}, \href{http://arxiv.org/abs/2004.00598}{{\ttfamily
  arXiv:2004.00598 [hep-th]}}.

\bibitem{Grumiller:2017qao}
D.~Grumiller, R.~McNees, J.~Salzer, C.~Valc\'arcel, and D.~Vassilevich,
  ``{Menagerie of AdS$_{2}$ boundary conditions},''
  \href{http://dx.doi.org/10.1007/JHEP10(2017)203}{{\em JHEP} {\bfseries 10}
  (2017) 203}, \href{http://arxiv.org/abs/1708.08471}{{\ttfamily
  arXiv:1708.08471 [hep-th]}}.

\bibitem{Grumiller:2007ju}
D.~Grumiller and R.~McNees, ``{Thermodynamics of black holes in two (and
  higher) dimensions},''
  \href{http://dx.doi.org/10.1088/1126-6708/2007/04/074}{{\em JHEP} {\bfseries
  04} (2007) 074}, \href{http://arxiv.org/abs/hep-th/0703230}{{\ttfamily
  arXiv:hep-th/0703230}}.

\bibitem{Ruzziconi:2020wrb}
R.~Ruzziconi and C.~Zwikel, ``{Conservation and Integrability in
  Lower-Dimensional Gravity},''
  \href{http://dx.doi.org/10.1007/JHEP04(2021)034}{{\em JHEP} {\bfseries 04}
  (2021) 034}, \href{http://arxiv.org/abs/2012.03961}{{\ttfamily
  arXiv:2012.03961 [hep-th]}}.

\bibitem{Harlow:2018tqv}
D.~Harlow and D.~Jafferis, ``{The Factorization Problem in Jackiw-Teitelboim
  Gravity},'' \href{http://dx.doi.org/10.1007/JHEP02(2020)177}{{\em JHEP}
  {\bfseries 02} (2020) 177}, \href{http://arxiv.org/abs/1804.01081}{{\ttfamily
  arXiv:1804.01081 [hep-th]}}.

\bibitem{Harlow:2015lma}
D.~Harlow, ``{Wormholes, Emergent Gauge Fields, and the Weak Gravity
  Conjecture},'' \href{http://dx.doi.org/10.1007/JHEP01(2016)122}{{\em JHEP}
  {\bfseries 01} (2016) 122}, \href{http://arxiv.org/abs/1510.07911}{{\ttfamily
  arXiv:1510.07911 [hep-th]}}.

\bibitem{Guica:2015zpf}
M.~Guica and D.~L. Jafferis, ``{On the construction of charged operators inside
  an eternal black hole},''
  \href{http://dx.doi.org/10.21468/SciPostPhys.3.2.016}{{\em SciPost Phys.}
  {\bfseries 3} no.~2, (2017) 016},
  \href{http://arxiv.org/abs/1511.05627}{{\ttfamily arXiv:1511.05627
  [hep-th]}}.

\bibitem{Gao:2021uro}
P.~Gao, D.~L. Jafferis, and D.~K. Kolchmeyer, ``{An effective matrix model for
  dynamical end of the world branes in Jackiw-Teitelboim gravity},''
  \href{http://arxiv.org/abs/2104.01184}{{\ttfamily arXiv:2104.01184
  [hep-th]}}.

\bibitem{Harlow:2020bee}
D.~Harlow and E.~Shaghoulian, ``{Global symmetry, Euclidean gravity, and the
  black hole information problem},''
  \href{http://dx.doi.org/10.1007/JHEP04(2021)175}{{\em JHEP} {\bfseries 04}
  (2021) 175}, \href{http://arxiv.org/abs/2010.10539}{{\ttfamily
  arXiv:2010.10539 [hep-th]}}.

\bibitem{Papadodimas:2012aq}
K.~Papadodimas and S.~Raju, ``{An Infalling Observer in AdS/CFT},''
  \href{http://dx.doi.org/10.1007/JHEP10(2013)212}{{\em JHEP} {\bfseries 10}
  (2013) 212}, \href{http://arxiv.org/abs/1211.6767}{{\ttfamily arXiv:1211.6767
  [hep-th]}}.

\bibitem{Marolf:2015jha}
D.~Marolf, ``{Comments on Microcausality, Chaos, and Gravitational
  Observables},'' \href{http://dx.doi.org/10.1088/0264-9381/32/24/245003}{{\em
  Class. Quant. Grav.} {\bfseries 32} no.~24, (2015) 245003},
  \href{http://arxiv.org/abs/1508.00939}{{\ttfamily arXiv:1508.00939 [gr-qc]}}.

\bibitem{Blake:2013owa}
M.~Blake, D.~Tong, and D.~Vegh, ``{Holographic Lattices Give the Graviton an
  Effective Mass},''
  \href{http://dx.doi.org/10.1103/PhysRevLett.112.071602}{{\em Phys. Rev.
  Lett.} {\bfseries 112} no.~7, (2014) 071602},
  \href{http://arxiv.org/abs/1310.3832}{{\ttfamily arXiv:1310.3832 [hep-th]}}.

\bibitem{Donnelly:2015hta}
W.~Donnelly and S.~B. Giddings, ``{Diffeomorphism-invariant observables and
  their nonlocal algebra},''
  \href{http://dx.doi.org/10.1103/PhysRevD.93.024030}{{\em Phys. Rev. D}
  {\bfseries 93} no.~2, (2016) 024030},
  \href{http://arxiv.org/abs/1507.07921}{{\ttfamily arXiv:1507.07921
  [hep-th]}}. [Erratum: Phys.Rev.D 94, 029903 (2016)].

\end{thebibliography}\endgroup

\end{document}